\documentclass{aa}  

\usepackage{graphicx}
\usepackage{txfonts}
\usepackage{subcaption}         
\usepackage{lscape}             
\usepackage{placeins}           
\usepackage{multirow}  
\usepackage{xcolor}      
\usepackage{amsfonts,amsmath,amssymb, mathrsfs}                        

\newcommand{\src}{GX~340$+$0}

\newcommand{\nicer}{\textit{NICER}}
\newcommand{\astr}{\textit{AstroSat}}
\newcommand{\ixpe}{\textit{IXPE}}
\newcommand{\nustar}{\textit{NuSTAR}}
\newcommand{\hxmt}{\textit{Insight}-HXMT}
\newcommand{\atca}{\textit{ATCA}}
\newcommand{\gmrt}{\textit{GMRT}}

\defcitealias{Bhargava2023ApJ...955..102B}{B23}

\begin{document}

   \title{X-ray and Radio campaign of the Z-source GX~340$+$0}

   \subtitle{Discovery of X-ray polarization and its implications}

%
%
%

   \author{ Yash Bhargava\inst{1,2}\corrauth{yash.bhargava@inaf.it}
          \and Mason Ng\inst{3,4,5}
          \and Liang Zhang\inst{6}
          \and Arvind Balasubramanian\inst{7}
          \and Thomas Russell\inst{8}
          \and Aman Kaushik\inst{2}
          \and Vishal Jadoliya\inst{9}
          \and Swati Ravi\inst{3}
          \and Sudip Bhattacharyya\inst{2}
          \and Mayukh Pahari\inst{9}
          \and Jeroen Homan\inst{10}
          \and Herman L. Marshall\inst{3}
          \and Deepto Chakrabarty\inst{3}
          \and Francesco Carotenuto\inst{11}
        }

   \institute{ INAF-Osservatorio Astronomico di Cagliari, via della Scienza, 09047, Selargius (CA), Italy   
   \and Department of Astronomy and Astrophysics, Tata Institute of Fundamental Research, 
1 Homi Bhabha Road, Colaba, Mumbai 400005, India
   \and MIT Kavli Institute for Astrophysics and Space Research, Massachusetts Institute of Technology, Cambridge, MA 02139, USA
   \and Department of Physics, McGill University, 3600 rue University, Montréal, QC H3A 2T8, Canada
   \and Trottier Space Institute, McGill University, 3550 rue University, Montréal, QC H3A 2A7, Canada
   \and Key Laboratory of Particle Astrophysics, Institute of High Energy Physics, Chinese Academy of Sciences, Beijing 100049, China
   \and Indian Institute of Astrophysics, Koramangala II Block, Bangalore 560034, India
   \and INAF, Istituto di Astrofisica Spaziale e Fisica Cosmica, Via U. La Malfa 153, I-90146 Palermo, Italy
   \and Department of Physics, Indian Institute of Technology Hyderabad, IITH main road, Kandi 502284
   \and Eureka Scientific, Inc., 2452 Delmer Street, Oakland, CA 94602, USA
   \and INAF-Osservatorio Astronomico di Roma, Via Frascati 33, I-00078, Monte Porzio Catone (RM), Italy
   }
   \date{Received September 30, 20XX}

  \abstract
{We present a study of X-ray polarization from the neutron star (NS) low-mass X-ray binary Z-source, GX~340$+$0, using an Imaging X-ray Polarimetry Explorer (\ixpe)
observation taken in March 2024. Along with the \ixpe\ observation, we conducted an extensive X-ray and radio monitoring campaign, tracing the source properties during and around the \ixpe\ observation.
The source appeared to be on the horizontal branch (HB) throughout the multi-wavelength campaign. 
We detected significant X-ray polarization($>11\sigma$) in 2--8~keV with a polarization degree (PD) of $4.02 \pm 0.35$\% and polarization angle (PA) = $37.6 \pm 2.5\degr$ ($1\sigma$ confidence interval). 
The polarization appears to be energy dependent, where the 2--2.5~keV PA is lower, $9\pm8\degr$, while the higher energy bands are consistent with the PA found over the 2--8~keV energy band ($\sim 37\degr$). 
The simultaneous \astr-\ixpe\ spectro-polarimetric observations suggest that the 2--8~keV energy band is dominated by the polarized emission from the Comptonized blackbody component with a marginal contribution from the accretion disk. 
Radio observations in the 0.7--9 GHz frequency range show a non-detection in the 0.7--1.5 GHz range, but the source was significantly detected in the 5.5--9 GHz band, suggesting the presence of a spectral break between 1.5 and 5.5~GHz. Our \atca\ observations do not detect any radio polarization, with a  3-$\sigma$ upper limit on the linear polarization of $<$6\% and $<$4\% on the circular polarization (centered at 7.25\,GHz). We discuss the origin of the X-ray polarization and its implications on the geometry of the spectral components. }

   \maketitle

\nolinenumbers
\section{Introduction}
Z-sources are bright, low-mass X-ray binaries (LMXB) with a neutron star (NS) primary accreting from a companion star. They show a unique `Z' shaped track in their X-ray hardness-intensity diagram (HID) and the color-color diagram \citep[CCD;][]{hasinger1989A&A...225...79H}, as they cycle through their spectral states. 
The track itself is divided into the horizontal branch (HB), normal branch (NB), and flaring branch (FB), with the `hard apex' joining the HB and the NB and the `soft apex' joining the NB and the FB. 
These sources host a weak magnetic field NS (typically $10^{8-9}$~G) and accrete at 
close to the Eddington limit (which depends on the NS mass and composition of the accreting plasma, e.g. \citealt{vdkReview2004astro.ph.10551V}). The transitions across different branches are smooth, always passing through the apexes.

The literature distinguishes two subclasses of the Z-sources, depending on the shape of the `Z' traced: Cyg-like sources (which include Cyg~X-2, \src, and GX~5$-$1) and Sco-like sources (which include Sco~X-1, GX~17$+$2, and GX~349$+$2). The evolution along the track is often linked to strong changes in the mass accretion rate \citep{homan2007ApJ...656..420H, lin2009ApJ...696.1257L, Homan2010ApJ...719..201H, Chakrabortyetal2011, 2024ApJ...966..232N}. The classes themselves are perhaps related to the mass accretion rate, as shown by the transient sources XTE J1701$-$462 and IGR J17480
$-$2446 which not only trace Z-tracks of both Sco-like and Cyg-like sources but also at lower accretion rates trace atoll source behavior \citep{homan2007ApJ...656..420H, lin2009ApJ...696.1257L, Homan2010ApJ...719..201H}.

The Z-track typically depicts secular shifts on the HID/CCD space. The stability of the shape of the Z-track indicates that there is a clear and specific spectral evolution as the source evolves. To understand the cause of such a spectral evolution, one needs to investigate the components involved in the emission. 
The persistent Z sources mentioned above have been extensively probed using the spectro-timing analysis, which yielded a crude picture with many degeneracies in the emission components \citep{done2007A&ARv..15....1D}. The emission in 2--20~keV energy range is well modeled as a combination of a thermal and a Comptonization component, with the thermal component being an accretion disk \citep[eastern model or EM;][]{mitsuda1989PASJ...41...97M}, a blackbody component \citep[western model or WM; ][]{Church2006A&A...460..233C} or a combination of two thermal components with harder non-thermal component \citep{lin2007ApJ...667.1073L, lin2009ApJ...696.1257L, lin2012ApJ...756...34L}. The Comptonized emission arises from the Compton up-scattering of photons by a hot plasma of electrons (also referred to as the corona; \citealt{mitsuda1989PASJ...41...97M, nthzdz1996MNRAS.283..193Z, Church2006A&A...460..233C, Seifina2013ApJ...766...63S,  Bhargava2023ApJ...955..102B}, hereafter \citetalias{Bhargava2023ApJ...955..102B}, \citealt{Putha2024MNRAS.532.3961P}). 
Some of the spectral modeling approaches have also inferred the presence of a transient hard tail (generally extending up to $\sim100$~keV) in the spectra, typically dependent on the position of the source in the Z-track evolution \citep[e.g.][]{paizis2006A&A...459..187P}. 
The investigation of Z-sources using soft X-ray instruments demonstrates the presence of a tail to the soft thermal component, which can be modeled as a blackbody \citep[e.g.][]{Seifina2013ApJ...766...63S} or disk emission \citepalias[e.g.][]{Bhargava2023ApJ...955..102B}.

The polarimetric observation of Z-sources with the Imaging X-ray Polarimetry Explorer (\ixpe) has already provided new insights into the orientation of various emission components.  
In the case of Cyg X-2, the polarization degree (PD) of $1.8\pm0.8$\% \citep{Farinelli2023MNRAS.519.3681F} was measured during the NB and $4.2\pm1.1$\% during the FB \citep{Gnarini2025A&A...699A.230G, 2026A&A...711A.208G}. The studies of Cyg X-2 indicates that the polarization from the accretion disk is marginal, but polarization angles (PA) of the emission from the disk and the corona may be perpendicular. The disk polarization, in such a case, could arise from the disk atmosphere \citep{Sunyaev1985A&A...143..374S} and would be perpendicular to the Comptonized component \citep{Farinelli2023MNRAS.519.3681F}. 
The transient Z-source XTE~J1701$-$462 demonstrated a variable X-ray polarization when observed during the HB and NB transition of the source \citep{Cocchi2023A&A...674L..10C}. During the HB, the source showed a PD of $4.6\pm0.4$\%, while during the NB, the PD dropped to $\sim0.6$\%, indicating that the polarization in Z-sources is indeed dependent on the Z-track position.    
GX~5$-$1 was observed in the HB and NB, and the PD in the HB ($4.3\pm0.3$\%) was observed to be significantly higher than that in the NB \citep[$2.0\pm0.3$\%;][]{Fabiani2024A&A...684A.137F}, depicting a dependence of the PD on the Z-track, similar to XTE~J1701$-$462. 
GX~5$-$1 also hinted towards a difference in the PA of the spectral components, but the dependence was not as distinct as observed in Cyg~X-2. 
Sco X-1 was observed during its NB and FB, showing a lower polarization \citep{monaca2024ApJ...960L..11L} when compared to the NB polarization in either Cyg~X-2 or GX~5$-$1, indicating that the degree of polarization may be dependent on the shape of the Z-track as well. Cir~X-1 (which also shows intermittent Z-source-like behavior) indicated a variation in the PA as a function of the X-ray hardness, indicating a clear interplay between the spectral components \citep{rankin2024ApJ...961L...8R}.
Additionally, spectro-polarimetric simulations for weakly magnetized sources predict a typical X-ray polarization of $\approx$4\% with the energy dependence of PD/PA breaking the degeneracies in the spectral decomposition of such sources \citep{gnarini2022MNRAS.514.2561G}. The observations have been consistent with the non-detections of the boundary layer polarization \citep{Farinelli2024A&A...684A..62F} and the surface layer polarization \citep{Bobrikova2025A&A...696A.181B} in weakly magnetized NS LMXBs. \citet{gnarini2024A&A...690A.230G}, have conducted simulations with modified geometries of EM and WM for a dipping atoll 4U 1624-49, which seems to prefer EM over WM to explain the observed polarization properties. Recently, \citet{Gnarini2025A&A...699A.230G} and \citet{2026A&A...711A.208G} compared the polarization properties of all known Z-sources uniformly, reaffirming the typical values of the PD, as well as the dependence of the polarization on the Z-track position. 

\src\ is one of the bright, persistent sources that demonstrates a complete trace of the Z-track \citep{mitsuda1989PASJ...41...97M, Jonker1998ApJ...499L.191J, Jonker2000ApJ...537..374J,Gilfanov2003A&A...410..217G,lavagetto2004NuPhS.132..616L,Iaria2006ChJAS...6a.257I,balucinska2010A&A...512A...9B,Seifina2013ApJ...766...63S}. 
The source traverses through all of the Z branches in a relatively short (a few days) timescale \citep{Jonker2000ApJ...537..374J}. The source is obscured at optical wavelengths but has been associated with a possible infrared counterpart \citep{Miller1993AJ....106...28M}. The source is a known radio emitter \citep{Penninx1993A&A...267...92P}, which appears to be strongly correlated to the X-ray flux \citep{oosterbroek1994A&A...281..803O, 2000MNRAS.318..599B}. 
The binary inclination of the source is unknown due to a lack of optical measurements, but since the source doesn't exhibit any dips (in contrast with Cyg X-2, e.g. \citealt{Mondal2018MNRAS.474.2064M}), the inclination is expected to be $\lesssim70^\circ$ \citep{frank87}. Reflection modeling of the Fe K$\alpha$ feature imply an inclination of the source to be $\approx37^\circ$ \citep{dai2009ApJ...693L...1D, miller2016ApJ...822L..18M, monaca2024A&A...691A.253L}. However, \citet{Seifina2013ApJ...766...63S} suggest that the inclination of the system is high (i.e., we view the system close to edge-on), and the Z-track evolution is caused by the changing size of the Comptonizing medium.

One of the important unanswered questions about this source is the nature of the corona and its geometry, whether it covers the disk \citepalias{Bhargava2023ApJ...955..102B}, is a puffed up optically thin medium close to the NS surface \citep{Seifina2013ApJ...766...63S}, it arises from the Comptonization from BL/SL \citep{Inogamov1999AstL...25..269I, Popham2001ApJ...547..355P}, or it aligns with the radio jet \citep{oosterbroek1994A&A...281..803O}. To understand the geometry of various components of the system, we conducted the first \ixpe\ observation of  \src\ supported by an extensive campaign using various X-ray and radio observatories. In this article, we report the results from an X-ray polarimetric, spectral, and timing study of the source, as well as report on supporting radio observations. 

\section{Observations and Data Reduction} \label{appsec:data_red}

\begin{figure}
  \centering
  \includegraphics[width=\columnwidth]{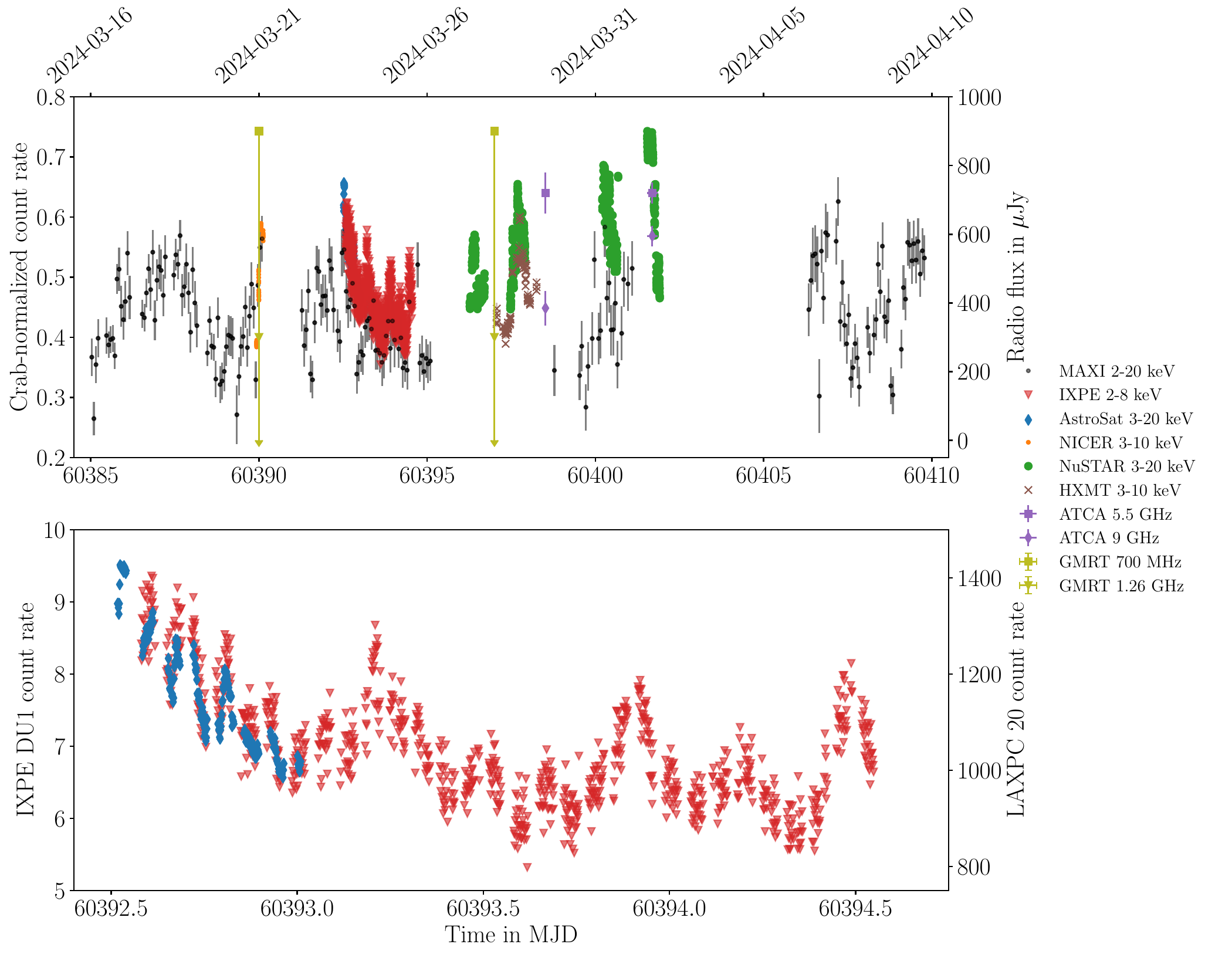}
  \caption{\textit{Top panel}: The orbit-wise Crab-normalized light curve from \textit{MAXI}/GSC over-plotted with light curves from various missions (indicated in different symbols and colors as per the legend) during and around the \ixpe\ observation. The light curves are background-subtracted and Crab-normalized except for the \ixpe\ light curve, which is scaled by a factor of 15.  The radio flux densities from \atca\ are indicated using the right axis, while \gmrt\ did not detect the source in both Band 4 (750~MHz) and Band 5 (1260~MHz), as such 3$\sigma$ upper limits are shown for both bands. \textit{Bottom panel}: The \ixpe\ DU1 observation light curve (in 2--8~keV) is over-plotted with simultaneous \astr-LAXPC 20 (in 3--20~keV) light curve. Both light curves are binned at 100~s. }\label{fig:full_lc}
\end{figure}

\ixpe\ observed the source during the latter half of March 2024 as detailed in Table~\ref{tab:obslog}. 
Since \src\ is variable on a much shorter timescale, it is paramount to have simultaneous observations with instruments with spectro-timing capabilities to identify the exact state and investigate the branch-resolved polarization. We coordinated a multi-wavelength campaign of the source to observe the source simultaneously or at least {close to \ixpe\ observations}. The details of these observations are noted in Table~\ref{tab:obslog}, and a combined light curve of the campaign is shown in Figure~\ref{fig:full_lc}.  To plot in the same scale, the X-ray light curves have been background-subtracted and normalized to the Crab count rates for each respective instrument, except for the \ixpe\ light curve, which is scaled by a factor of 15 (the reason for scaling the \ixpe\ observation is described in section~\ref{appsec:data_red:ixpe}) The details of the data reduction for individual observatories and instruments are mentioned in various subsections of section~\ref{appsec:data_red}.

\begin{table*}
  \centering
  \caption{Observation log for the multi-wavelength campaign of \src}\label{tab:obslog}
  \resizebox{1.7\columnwidth}{!}{
  \begin{tabular}{|c|llc|ccc|}
    \hline
    Epoch & Observatory & Instrument & Observation ID & Start Time & Stop Time & Exposure (in ks) \\ \hline
    \multirow{2}{*}{0} & \nicer & XTI & 7108010101-2 & 2024-03-20 22:14:20 & 2024-03-21 03:01:08 & 2.4 \\
     & \gmrt & -- & ddtC332 & 2024-03-20 22:30:00 & 2024-03-21 01:30:00 & 3.54\\ \hline
    \multirow{3}{*}{1} & \ixpe & GPD (DU1--3) & 03003301 & 2024-03-23 13:56:24 & 2024-03-25 13:09:44 & 99.0\\
     & \multirow{2}{*}{\astr} & SXT & \multirow{2}{*}{9000006138} & 2024-03-23 12:23:48 & 2024-03-24 00:15:28 & 14.2 \\
     &  & LAXPC & & 2024-03-23 12:24:50 & 2024-03-24 00:13:10 & 23.1  \\ \hline
    \multirow{6}{*}{2}  & \nustar & FPMA/B & 91002313002 & 2024-03-27 06:24:26 & 2024-03-27 16:53:46 & 12.4\\
     & \multirow{2}{*}{\hxmt} & LE &P061437700101- & 2024-03-28 01:33:25 & 2024-03-29 06:04:41 & 6.46\\
     &  & ME  & P061437700109& 2024-03-28 01:32:54 & 2024-03-29 06:53:30 & 13.98\\
     & \gmrt  & -- & ddtC332 & 2024-03-27 22:30:00 & 2024-03-28 01:30:00 & 3.54\\ 
     & \nustar & FPMA/B & 91002313004 & 2024-03-28 11:15:40 & 2024-03-28 21:45:00 & 12.5 \\
     & \atca & -- & CX566 & 2024-03-29 11:29:40 & 2024-03-29 13:18:20 & 6.0\\ \hline
    \multirow{3}{*}{2+} & \nustar & FPMA/B &91002313006 & 2024-03-31 04:57:17 &  2024-03-31 16:13:33 & 11.4\\
     & \nustar & FPMA/B & 91002313008& 2024-04-01 13:00:41 & 2024-04-01 21:54:01 & 10.9 \\
     & \atca & -- & CX566 & 2024-04-01 12:43:20 & 2024-04-01 20:25:30 & 14.04\\
    \hline   

  \end{tabular}
  }
\end{table*}

\subsection{IXPE} \label{appsec:data_red:ixpe}
\ixpe\ \citep{2022JATIS...8b6002W} observed \src\ for an exposure of $\approx99$~ks from 2024-03-23 13:56:24 (UTC) to 2024-03-25 13:09:44 (UTC) using all three gas pixel detector units (referred to as DU1, DU2, and DU3). Due to a spacecraft anomaly, some of the engineering data could not be recovered, and the data was estimated by extrapolating from the rest of the data. Hence, for the observation, special processing was conducted by the \ixpe\ Science Operation Center (SOC) team, and the level 2 files thus produced were utilized for the analysis. 

We used \texttt{ixpeobssim} software (v31.0.1) to process the observations and extract the polarization properties of the source \citep{2015APh....68...45K,ixpeobssim2022SoftX..1901194B}. We selected the source events from a circular region of radius 1.6$\arcmin$ and the background region from an annular region of inner radius of 2.5$\arcmin$ and outer radius of 5$\arcmin$. We binned the selected events using different algorithms, viz. \textsc{pcube}, \textsc{phaq}, \textsc{phau} to generate polarization cubes, Q spectra, and U spectra, respectively, for each DU. The calibration files version v13 \texttt{obssim20240101\_alpha075} were used to extract appropriately weighted data products \citep{diMarco2022AJ....163..170D}. For the model-independent analysis, the polarization information from the detectors was combined and depicted in Figure~\ref{fig:pcube}. 

To plot the \ixpe\ light curve on the same scale as all instruments in Figure~\ref{fig:full_lc}, we scaled the \ixpe\ count rates by a factor of 15. \src\ appeared as a point source in the \ixpe\ image. However, the Crab observation (to calibrate the count rate to the Crab) showed structure from the pulsar wind nebula. This caused a strong dependence of the count rate on the region size, and using a region of 1.6$\arcmin$ in Crab yielded a significantly higher count rate, resulting in a lower Crab-normalized count rate for \ixpe. Since the purpose of the figure is to show the simultaneous evolution of the source, we scaled the light curve arbitrarily by a factor of 15 to match the typical Crab normalized count rate as observed with other instruments. We depicted the actual count rate of the source in the bottom panel of Figure~\ref{fig:full_lc}.  
\subsection{NICER}

\nicer\ \citep{Gendreau2016SPIE.9905E..1HG} observed \src\ simultaneously with the first epoch of \gmrt\ observation (i.e., 2024 March 20-21) to characterize its X-ray state. The source was observed for 4 segments of $\sim700$~s each per International Space Station orbit ($\sim90$~min). The data were processed using the standard filtering criteria: a) excluding detectors 14 and 34, b) excluding South Atlantic Anomaly passages, c) considering the elevation angle for the bright Earth limb $>30^\circ$, and d) considering undershoot and overshoot rates (c/s/focal plane module) of 0--500 and 0--30, respectively,  with the tool \texttt{nicerl2} distributed as part of \textsc{nicerdas 2024-02-09\_V012} (\textsc{heasoft} v6.33). The light curves and spectra were extracted using the \texttt{nicerl3-lc} and \texttt{nicerl3-spect} with the background model \texttt{scorpeon} \citep{SCORPEON2024HEAD...2110536M}, which also allows for estimation of the background count rate in various energy bands. A similar analysis was conducted on a Crab observation from 2023 December 21 (observation ID: 6013010117) to normalize the count rates to Crab units.

\subsection{AstroSat}
Coordinating with the epoch of \ixpe\ observation, \astr\ \citep{Singh2014SPIE.9144E..1SS}  observed the source with Large Area X-ray Proportional Counter \citep[LAXPC;][]{Yadav2016SPIE.9905E..1DY,Yadav2017CSci..113..591Y} as a primary instrument and Soft X-ray telescope \citep[SXT;][]{singh2016SPIE.9905E..1ES, Singh2017JApA...38...29S} simultaneously. The \astr\ observation covered a total duration of 45~ks with a net exposure of 23.1 ks in LAXPC and 14.2 ks in SXT. The details of data reduction procedures for each instrument are summarized in the following sub-sections. 

\subsubsection{LAXPC} 
The orbit-wise level 1 data was downloaded from AstroBrowse\footnote{\url{https://astrobrowse.issdc.gov.in/astro_archive/archive/Home.jsp}} and the data was processed using the \textsc{LAXPCsoftware22Aug15}\footnote{\url{http://astrosat-ssc.iucaa.in/uploads/laxpc/LAXPCsoftware22Aug15.zip}} \citep{Antia2021JApA...42...32A,Misra2021JApA...42...55M}. 
\textsc{LAXPCsoftware22Aug15} includes the relevant calibration files, responses, and software to generate higher-level data products, including the merged event file for the full observation (\texttt{laxpc\_make\_event}). 
The intervals corresponding to Earth occultation and South Atlantic anomaly passages were excluded using the routine \texttt{laxpc\_make\_stdgti}. Using the built-in routines, the energy-dependent light curves, spectra, associated responses, background spectrum, and background light curves were extracted. For the analysis, we consider the data from the top layer (layer 1) due to the minimal background in the 3--20~keV band, limiting the spectral analysis to the same energy range. For the analysis, only the data from LAXPC 20 were considered, as the other two units were not suitable for spectral analysis (LAXPC 10 has shown abnormal gain, and LAXPC 30 suffered a gas leak early in the mission). For the spectral analysis of the data, we assume a 3\% systematic error (in line with previous studies, e.g., \citealt{Bhargava2019MNRAS.488..720B}, \citetalias{Bhargava2023ApJ...955..102B}, etc.). 
\subsubsection{SXT}

The orbit-wise level 2 processed data were accessed from AstroBrowse\footnote{\url{https://astrobrowse.issdc.gov.in/astro_archive/archive/Home.jsp}}, which has been processed with the latest version of the SXT pipeline (AS1SXTLevel2-1.5). 
Individual orbits were merged using the \textsc{Julia} tool \texttt{SXTMerger} as suggested by the SXT Payload Operation Center (POC). The standard products (i.e., spectra and light curves) were extracted using \texttt{xselect} (v2.5b from \textsc{heasoft} v6.32.1) for a circular region of 14$\arcmin$. For the extraction of the spectral products, only grade 0 events were selected, and the appropriate response matrix was obtained from the POC. The standard ancillary response file was modified for the selected region using the \texttt{sxtARFModule} tool provided by the POC. For the spectral analysis, we assume a 1.5\% systematic error (e.g. \citealt{sridhar2019MNRAS.487.4221S}, \citetalias{Bhargava2023ApJ...955..102B} etc.).

\subsection{NuSTAR}
\nustar\ \citep{Harrison2013ApJ...770..103H} carried out four Target of Opportunity (ToO) observations of \src\ on 2024 March 27, 28, 31, and April 1, for a total of 47.3~ks of filtered exposure (details of the observations are given in Table~\ref{tab:obslog}). The data were processed using the standard \nustar\ Data Analysis Software (\textsc{NuSTARDAS}) version 2.1.2 available under \textsc{HEASoft} 6.33, and with the CALDB version 20240405. We extracted the source events with a circular region of 120$\arcsec$ in radius with the centroid coordinates RA = 16h 45m 47.7s, DEC = -45$^\circ$ 36$\arcmin$ 40$\farcs$0. We also defined a background region in a source-free region with a circular region of radius 120$\arcsec$ with the centroid coordinates RA = 16h 46m 20.1184s, DEC = -45$^\circ$ 29$\arcmin$ 33$\farcs$231. We ran the \texttt{nupipeline} task to process the data and \texttt{nuproducts} to generate light curves and spectra. The light curves were binned with 128~s bins, and the spectra from the focal plane modules A and B (denoted as FPMA/FPMB) were regrouped to have a minimum of 30~counts per spectral bin. To facilitate comparison between the different instruments, we generated Crab-normalized light curves in different energy bands. We utilized the observation from 2024 March 18 (ObsID 11002303004, exposure=4.983~ks) and extracted the source region with centroid coordinates RA = 5h 34m 30.9s, DEC = +22$^\circ$ 00$\arcmin$ 53$\farcs$0 with a 200$\arcsec$ radius, and a background region with centroid coordinates RA = 5h 34m 21.1461s, DEC = +21$^\circ$ 52$\arcmin$ 55$\farcs$149, with a 150$\arcsec$ radius.
\subsection{Insight-HXMT}
\hxmt\ \citep{Zhang2020SCPMA..6349502Z} observed \src\ from 2024-03-28 01:21:40 to 2024-03-29 07-28:35. The data are extracted from all three instruments using the \hxmt\ Data Analysis software (HXMTDAS) v2.06\footnote{The data analysis software is available from \url{http://hxmten.ihep.ac.cn/software.jhtml}.}, 
and filtered with the following standard criteria: (1) pointing offset angle less than $0.04^{\circ}$; (2) Earth elevation angle larger than $10^{\circ}$; (3) the value of the geomagnetic cutoff rigidity larger than 8 GV;(4) at least 300 s before and after the South Atlantic Anomaly passage. To avoid possible contamination from bright Earth and nearby sources, we only use data from the small field of view (FoV) detectors \citep{Chen2018ApJ...864L..30C,Yang2022ApJ...937...33Y}.

\subsection{GMRT}

We observed \src\ with the wideband receiver backend of GMRT in two frequency bands - frequency band 4 (central frequency 750~MHz, bandwidth 400~MHz) and band 5 (central frequency 1260~MHz, bandwidth 400~MHz) taken on two epochs, 2024 March 20-21  and 2024 March 27-28  (Proposal ddtC332). The raw data were downloaded in the \texttt{FITS} format and converted to a \texttt{CASA} \citep{CASA2022PASP..134k4501C} measurement set format. The data were then calibrated (3C286 was used for flux calibration, and J1717-398 was used for phase calibration) and imaged using the automated continuum imaging pipeline \texttt{CASA-CAPTURE} \citep{Kale2021ExA....51...95K}. Eight rounds of self-calibration were done within each pipeline run. No sources were detected at the position of \src\ at either frequency band for both epochs. So, we place 3$\sigma$ upper limits of 900~$\mu$Jy in band 4 ($\approx$750~MHz) and 300~$\mu$Jy in band 5 ($\approx$1.3~GHz) by taking the RMS flux density of a large source-free region in the final image.

\subsection{ATCA}
\src\ was observed by the Australia Telescope Compact Array (\atca) on 2024-03-29 and 2024-04-01 under project code CX566. Observations on 2024-03-29 ran between 11:29:40 UT and 13:18:20 UT, while on 2024-04-01, \src\ was observed between 12:43:20 UT and 20:25:30 UT. Data were recorded at central frequencies of 5.5~GHz and 9~GHz, with 2~GHz of bandwidth at each frequency. For both epochs, we used PKS B1934$-$638 for primary bandpass and flux calibration and J1631$-$4345 for gain calibration. Data were calibrated and imaged, following standard procedures with the Common Astronomy Software Application \citep[\texttt{CASA} version 5.1.2;][]{CASA2022PASP..134k4501C}. Imaging used a Briggs robust parameter of zero, balancing sensitivity and angular resolution, as well as suppressing sidelobes from bright sources in the field.
\src\ was detected on both epochs. Flux densities (see Section 3.6) were measured by fitting for a point source in the image plane.

Our \atca\ observation taken on 2024-04-01 had sufficient hour angle coverage that we can measure the radio polarization. To do so, we used the unpolarised PKS B1934$-$638 to solve for antenna leakages and the \texttt{CASA} task \textit{qufromgains} to model the polarization angle of the gain calibrator. No linear (or circular) polarization was detected at either frequency. Stacking the 5.5~GHz and 9~GHz data together to provide our deepest limits gives 3$\sigma$ upper limits of $<$6\% on linear polarization and $<$4\% on the circular polarization (at 7.25~GHz).


\section{Data analysis \& Results} \label{sec:analysis}

\subsection{Model-independent X-ray polarimetric analysis}\label{ssec:pcube}

\begin{figure}
  \centering
  \includegraphics[width=0.7\columnwidth]{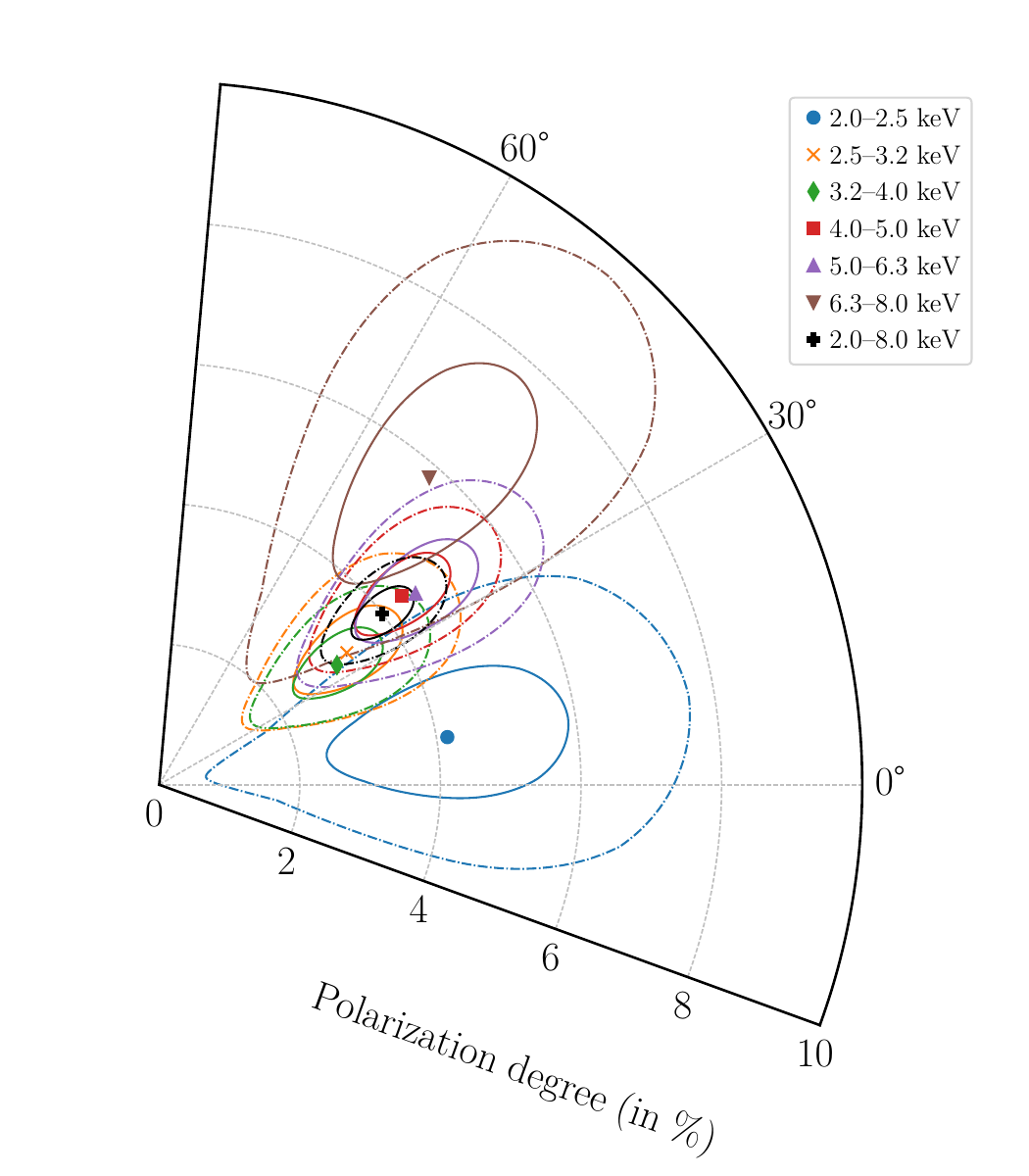} 
  \caption{Polar plot of the model-independent polarization of \src\ measured over various energy bands (see section~\ref{ssec:pcube}) using \textsc{pcube} analysis. The marker denotes the observed PD in radial coordinates and PA in azimuthal coordinates, while the ellipses show 1$\sigma$ \& 3$\sigma$ confidence intervals for each measurement. The colors correspond to different energy bands. The PA of 2--2.5~keV emission is marginally different from the rest of the energy bands. 
  }\label{fig:pcube}
\end{figure}

We investigated the polarimetric properties of the source by extracting the \textsc{pcube}  using \texttt{ixpeobssim} in various energy bands from the filtered events after appropriate weightings were applied (see section~\ref{appsec:data_red:ixpe} for details). \textsc{pcube} encapsulates the normalized Stokes Q and U information as a function of energy.  In the full energy band (2--8~keV), we detected significantly polarized emission from \src\ with PD = $4.02 \pm 0.35$\%   and PA = $37.6 \pm 2.5^\circ$ (black point in the  Figure~\ref{fig:pcube}). The confidence intervals reported in the text correspond to $1\sigma$ intervals unless indicated otherwise. In a model-independent fashion, we probed the energy dependence of the polarization by extracting the \textsc{pcube} in six logarithmically spaced bins over the 2--8~keV band  (depicted in the  Figure~\ref{fig:pcube} and noted in Table~\ref{tab:pol}). Most of the energy bands have a similar PA as the full 2--8~keV band, with the exception of 2--2.5~keV, which has a slightly lower PA of $9 \pm 8^\circ$ (although the PD was similar). We note that the larger error on the PA in the 2--2.5~keV energy band is due to a lower modulation fraction of individual detector units \citep{dimarco2022AJ....164..103D}, where similar uncertainties have been seen in previous \ixpe\ observations of other sources \citep[e.g.][]{rawat2023ApJ...949L..43R}. The PD of different energy bands does hint at a slight increase towards higher energies, but is consistent within the $2\sigma$ confidence level. We also validated the \textsc{pcube} results from the model-independent analysis with the \textsc{ixpe protractor} tool\footnote{\url{https://heasarc.gsfc.nasa.gov/docs/ixpe/analysis/contributed/ixpe_protractor.html}} and found that the results are consistent within the $1\sigma$ uncertainties. We have also included the \textsc{ixpe protractor} results in the Table~\ref{tab:pol}.

\begin{table}
    \centering
        \caption{Model-independent polarization properties of \src\ using \textsc{pcube} and \textsc{protractor} tools. The 1$\sigma$ confidence intervals are reported.}     \label{tab:pol}
      \resizebox{\columnwidth}{!}{
        \begin{tabular}{l|cc|cc}
         \multicolumn{5}{c}{Energy resolved  analysis} \\ \hline 
         & \multicolumn{2}{c|}{\textsc{pcube}} & \multicolumn{2}{c}{\textsc{protractor}} \\ \hline        
         Energy band (keV) & PD (\%) & PA ($^\circ$) & PD (\%) & PA ($^\circ$)  \\ \hline
         2.0--8.0 & $4.02\pm0.35$ & $37.5\pm2.5$ & $4.02 \pm0.28$  &  $36.1 \pm 2.0$ \\
         2.0--2.5 & $4.3\pm1.1$ &  $9\pm8$ & $4.4 \pm 1.2$  &  $ 7 \pm8$\\
         2.5--3.2 & $3.2\pm0.6$ & $35\pm5$ & $2.8 \pm 0.6$  &  $32 \pm6$\\
         3.2--4.0 & $3.0\pm0.5$ & $34\pm5$ & $3.3 \pm 0.5$  &  $32 \pm4$\\
         4.0--5.0 & $4.4\pm0.5$ & $38\pm3$ & $4.3 \pm 0.5$  &  $38 \pm3$ \\
         5.0--6.3 & $4.5\pm0.7$ & $37\pm4$ & $4.3 \pm 0.7$  &  $37 \pm4$\\
         6.3--8.0 & $5.8\pm1.2$ & $48\pm6$ & $5.7 \pm 1.0$  &  $50 \pm5$ \\ \hline \hline
         \multicolumn{5}{c}{Count rate resolved in 2--8~keV} \\ \hline
         & \multicolumn{2}{c|}{\textsc{pcube}} & \multicolumn{2}{c}{\textsc{protractor}} \\ \hline
         Count rate range & PD (\%) & PA ($^\circ$) & PD (\%) & PA ($^\circ$) \\ \hline
          5.1--6.8& $3.7\pm0.5$ & $36\pm4$ & $3.7 \pm 0.4$  &  $35.6 \pm  3.3$ \\ 
          6.8--9.5& $4.4\pm0.5$ & $38\pm3$ & $4.3 \pm 0.4$  &  $36.7 \pm  2.6$\\ \hline

    \end{tabular}
      }
\end{table}

We also investigated the polarization properties of the source as a function of the count rate (and by extension as a function of the position on the HB) by dividing the \ixpe\ observation into two subsets (high and low), using the median count rate ($\approx6.8$ counts/s in DU1) as the cut-off between the two subsets. The 2--8~keV polarization for these subsets (PD/PA for high: $4.35\pm0.47$\%/$38\pm3^\circ$ and for low: $3.69\pm0.53$\%$/36\pm4^\circ$) are consistent with each other within a 2$\sigma$ confidence interval (also noted in Table~\ref{tab:pol}). 

\subsection{Identification of the Z branch}

\begin{figure}
  \centering
\includegraphics[width=0.65\columnwidth]{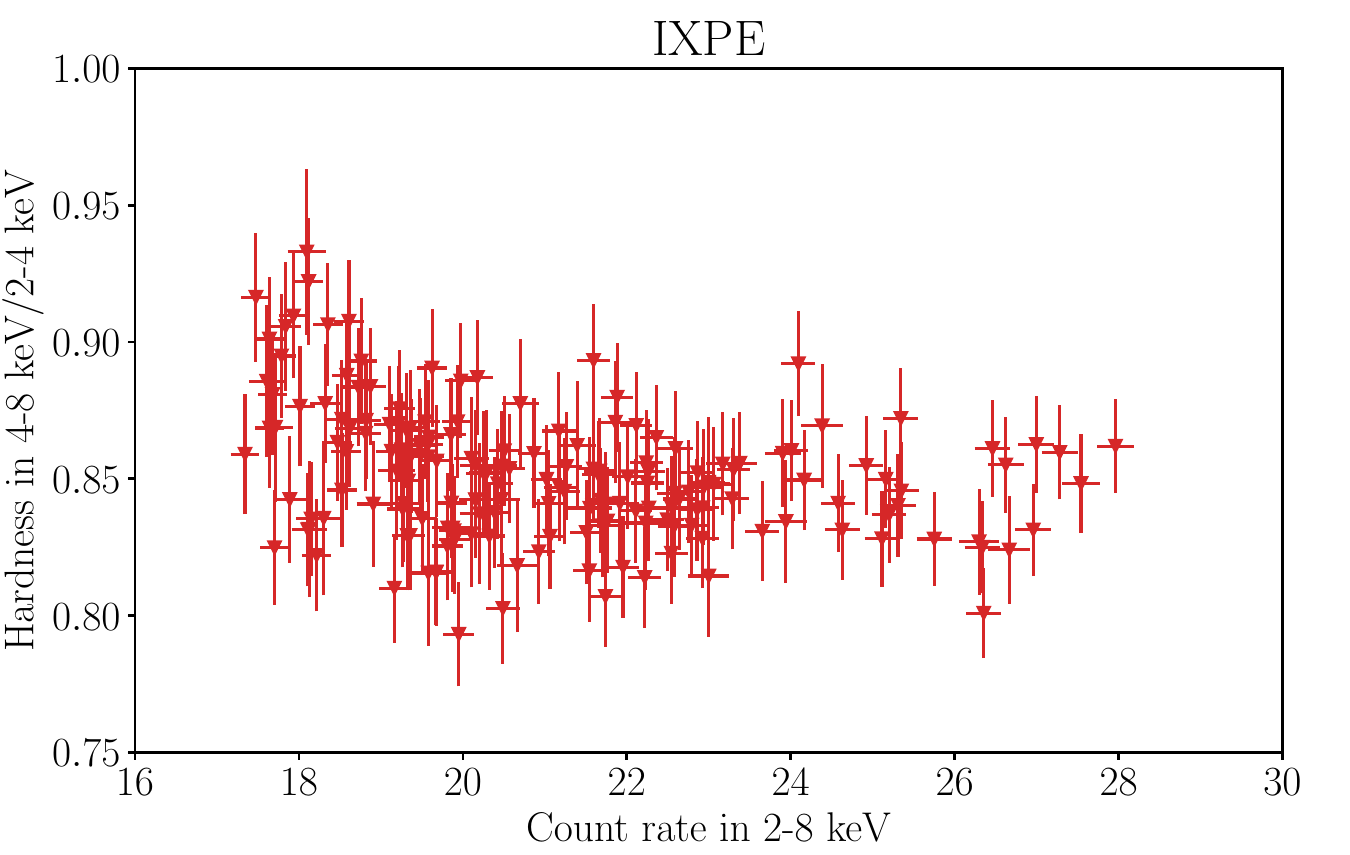} 
\includegraphics[width=0.65\columnwidth]{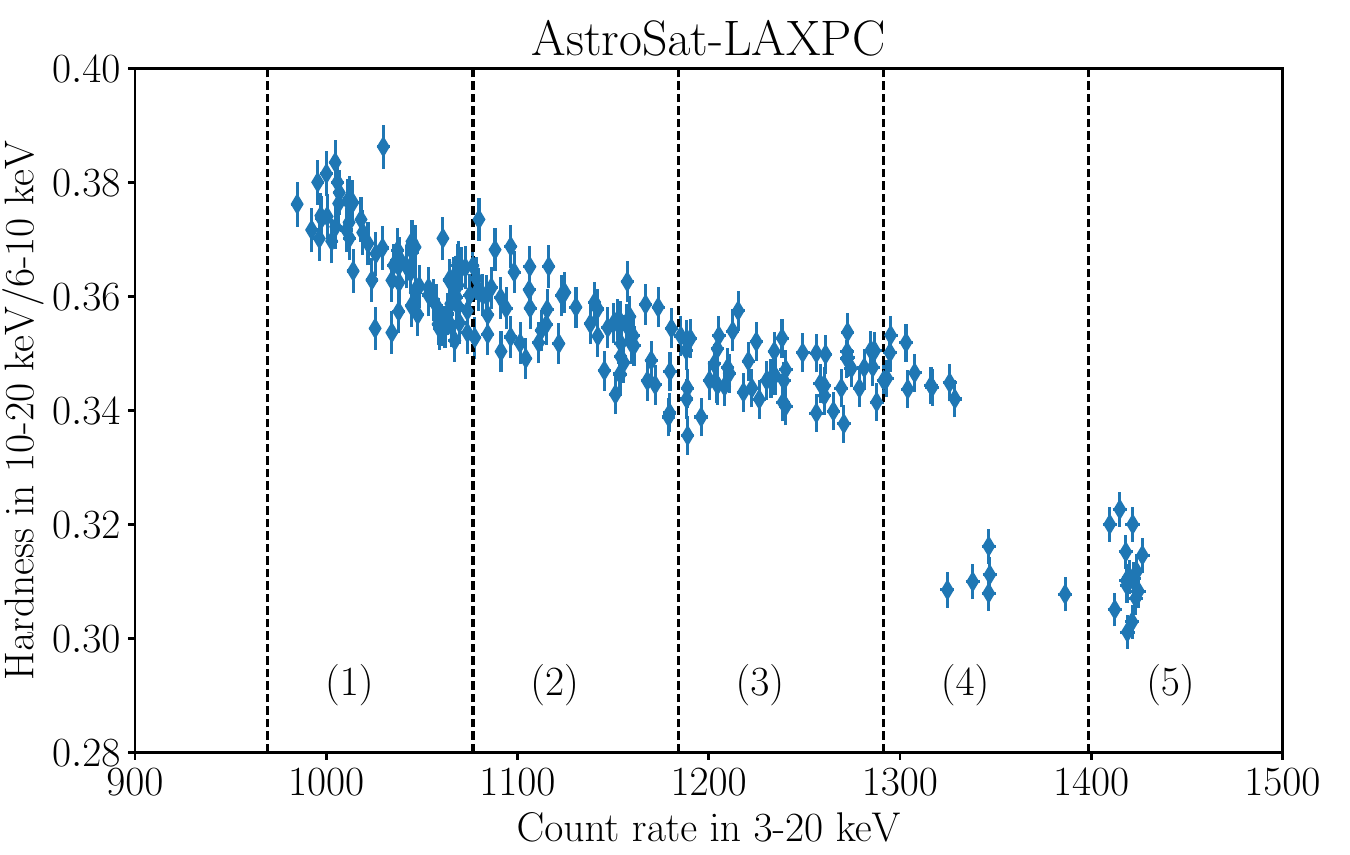} 
\includegraphics[width=0.65\columnwidth]{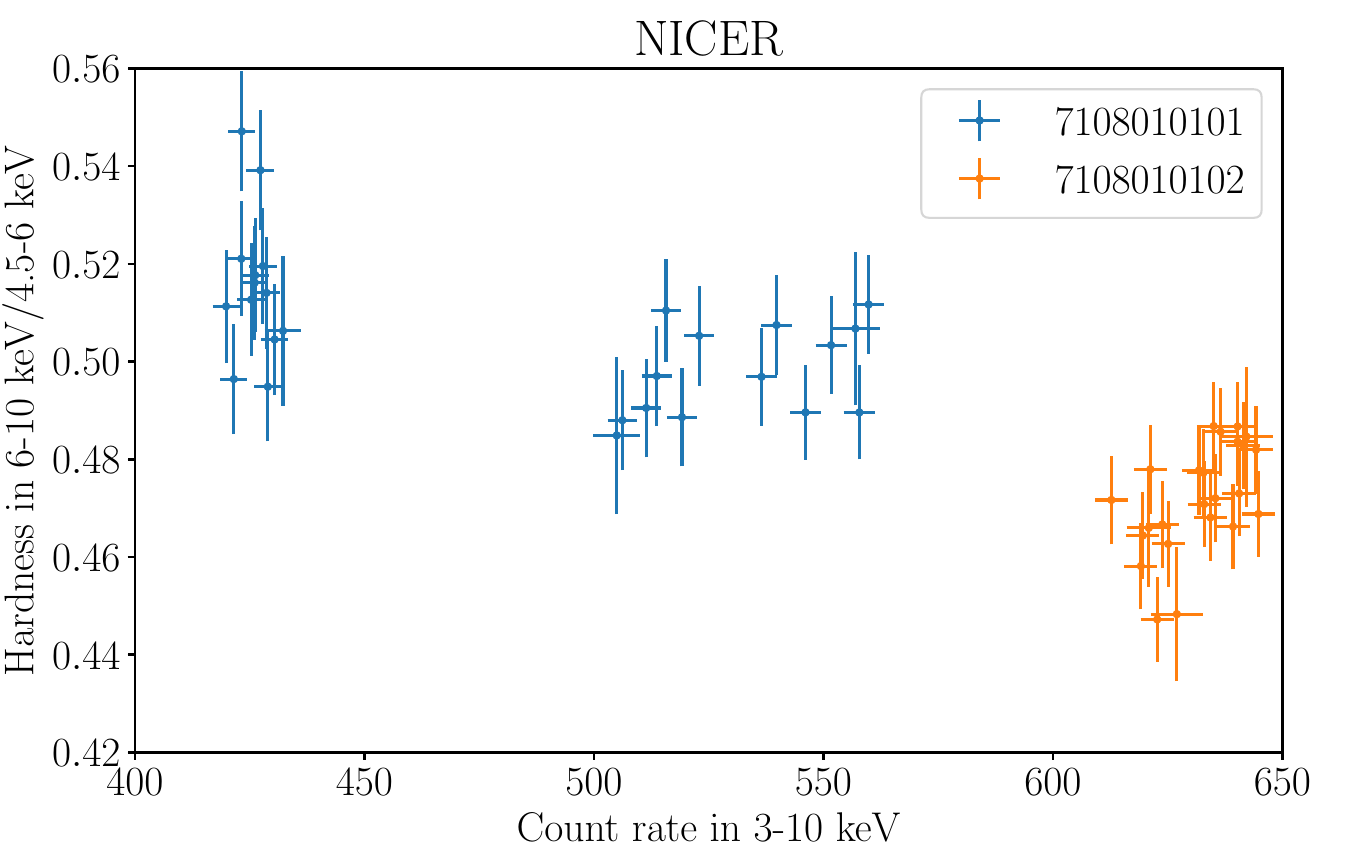} 
\includegraphics[width=0.65\columnwidth]{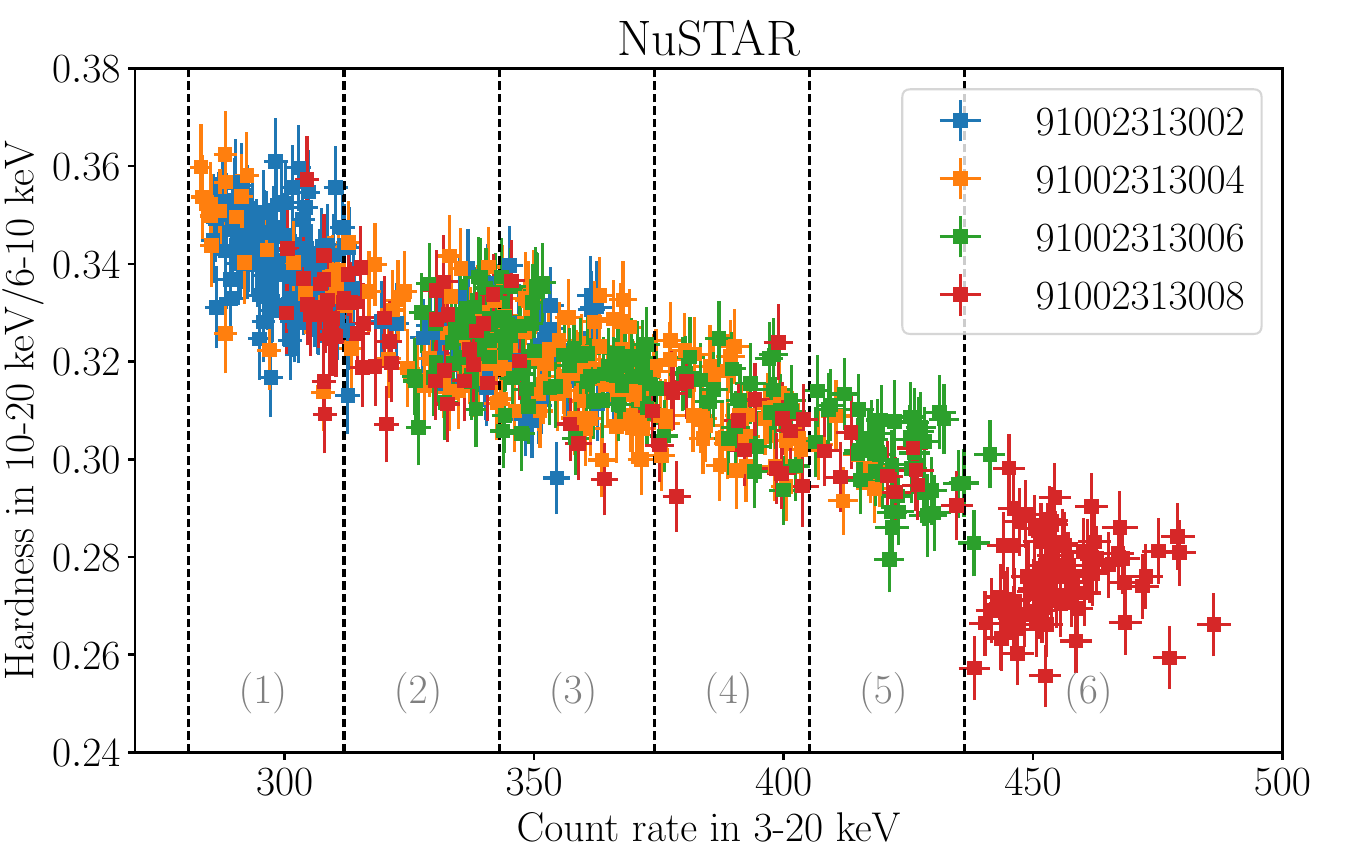} 
\includegraphics[width=0.65\columnwidth]{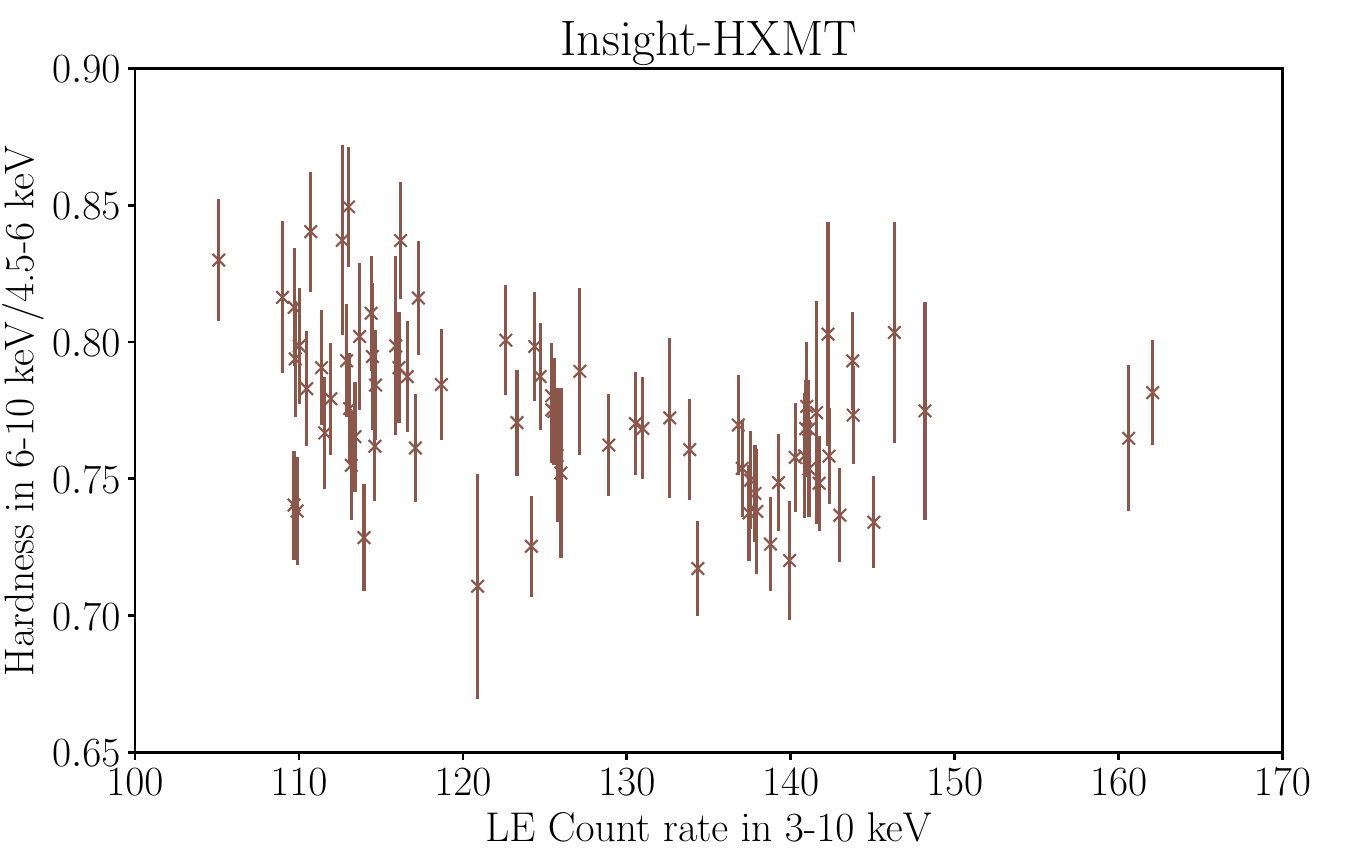} 
  
\caption{X-ray hardness-intensity diagrams (HIDs) from various observations of \src\ during the campaign in March 2024. Due to varied instrument energy coverage, we have considered different energy sub-bands for hardness ratio and total count rate computation. To improve the statistics of \ixpe\ HID, we have combined the counts from all the DUs. Comparing these with the typical HID of \src\ \citep[e.g.][]{Jonker2000ApJ...537..374J, Pahari2024MNRAS.528.4125P} and the timing properties (see Figure~\ref{fig:dps} and Table~\ref{tab:z-resolved-pds}), we can infer that the source was in the HB during these observations. Dashed vertical lines are marked in \astr\ and \nustar\ HIDs to indicate the zone boundaries for HID-resolved timing analysis.   }
  \label{fig:hid}
\end{figure}

\src\ shows a strong evolution of the spectro-timing properties as a function of the Z-track position. Thus, to compare the properties of the source across various observations, we constructed the HIDs of the source over various energy bands for different X-ray instruments (Figure~\ref{fig:hid}). In all the cases, the source was observed to follow a track along the HB branch \citep[see][for example]{Jonker2000ApJ...537..374J,Pahari2024MNRAS.528.4125P}. We verify the nature of the branch by investigating the timing properties of the source, in particular the 3--30~keV power density spectrum (PDS) in from \astr-LAXPC and the 3--30~keV co-spectrum from \nustar-FPMA/B. 
The presence of the quasi-periodic feature in 20--42~Hz (see section~\ref{ssec:timing} and Table~\ref{tab:z-resolved-pds} for more details) is a clear indication that the source was on the HB (\citealt{Jonker2000ApJ...537..374J}, \citetalias{Bhargava2023ApJ...955..102B}, \citealt{ Pahari2024MNRAS.528.4125P}).  
The \ixpe\ HID (figure~\ref{fig:hid}) indicates no change in the hardness with the count rate, suggesting that the source remained in the HB throughout this \ixpe\ observation. We do note that our second \ixpe\ observation (reported in \citealt{Bhargava2024arXiv241100350B}) identifies a clear transition to NB.

\subsection{Timing analysis}\label{ssec:timing}

\begin{figure}
  \centering
  \includegraphics[width=\columnwidth]{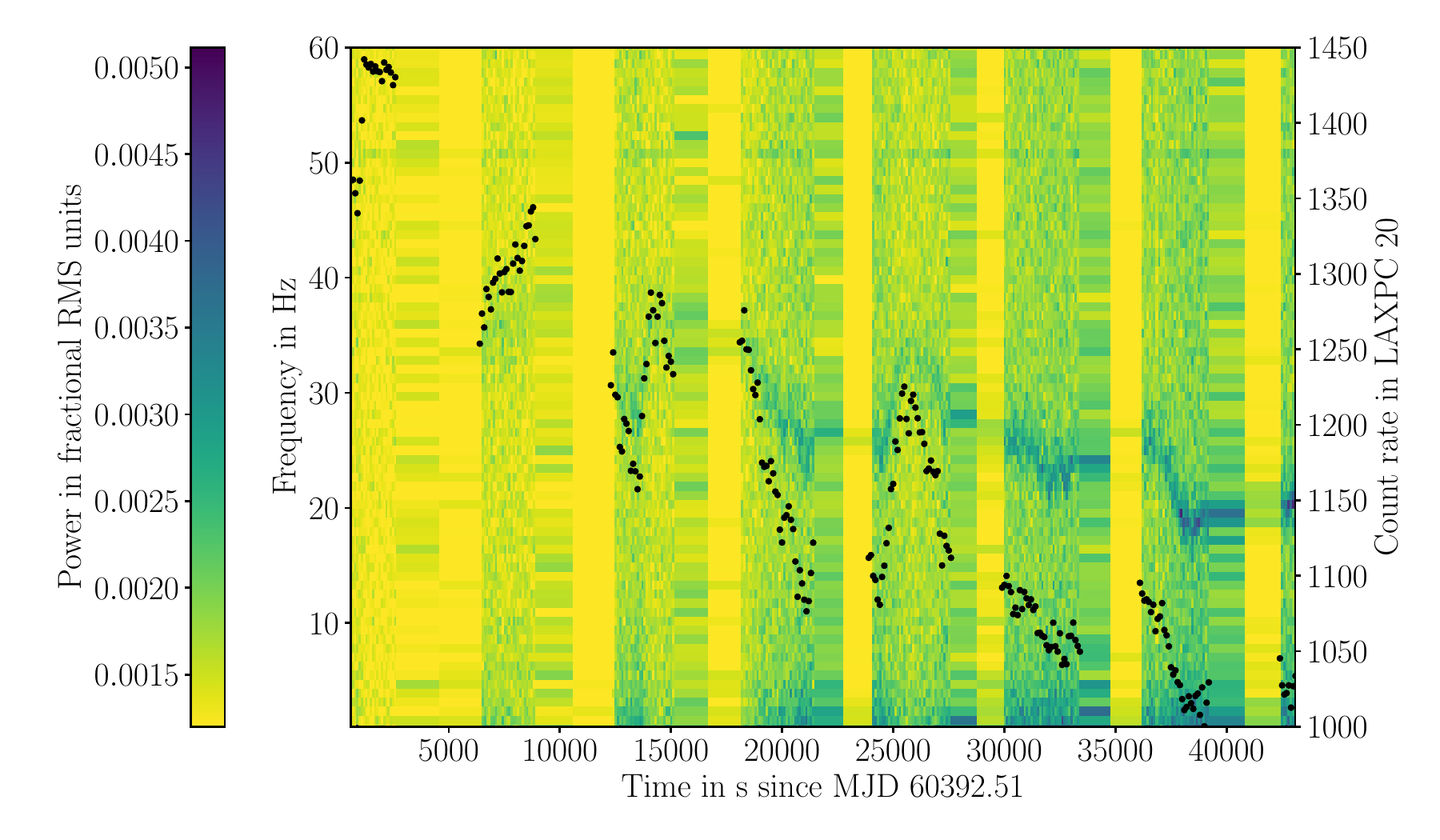}
  \caption{Dynamical power spectrum of \astr-LAXPC observation of \src\ during the March 2024 observation. The HBO is clearly observed varying from 30 Hz to 20 Hz during the observation. The black points trace the count rate evolution of the source (the count rates can be read on the right-hand axis), highlighting the correlation between the position of the source on the Z-track and the HBO frequency \citep[e.g.][]{Jonker2000ApJ...537..374J, vdkReview2004astro.ph.10551V}.  
  The segments where black points are absent correspond to gaps in the data. }
  \label{fig:dps}
\end{figure}

To confirm that the observation was conducted within the HB, we extracted the dynamical power spectrum (DPS) using the \astr-LAXPC data. We extracted a power spectrum from the detected events in the 3--30~keV band in 100~s long time slices between the frequencies of 1--100~Hz and depict the evolution of the power in various frequencies in Figure~\ref{fig:dps}. The DPS clearly highlights the presence of the low-frequency broadband noise and an evolving oscillation at 20--30 Hz (referred to as HB oscillation or HBO), indicative of the source being in the HB.

To compare the timing properties of the source during \astr\ and \nustar\ observations, we divided the observations into HID resolved zones. 
We normalized the count rates and the hardness to Crab units and then divided the \nustar\ crab-normalized HID into 6 equally spaced zones based on the Crab-normalized count rates. These intervals are marked in the figure~\ref{fig:hid} after rescaling the count rates to the observed values. Corresponding intervals are also plotted in the \astr\ HID for direct comparison (after rescaling the Crab normalized intervals back to observed values), but we note that the \astr\ observations only span zones 1-5, while the fourth \nustar\ observation also spans zone 6. 

From the \astr-LAXPC observation, we have extracted the rms-normalized, and Poisson-noise subtracted power density spectrum (PDS) for each HID zone in the 3--30~keV energy range after removing data gaps and count rate drops due to telemetry losses. Each PDS is fitted separately using a combination of power-law and multiple Lorentzians in the frequency range of 0.05--100~Hz. From the best-fit model components, total fractional rms, QPO frequencies, and their fractional rms were calculated. Fitted parameter values for the PDS corresponding to different zones are provided in Table~\ref{tab:z-resolved-pds}. The extent of zone 3 is broad, and thus, a double-peaked feature was detected in PDS. These features correspond to a single HBO, which shows a strong evolution (see figure~\ref{fig:dps} between the times 12000-15000~s).

Similarly, we analyzed the 3--30~keV \nustar\ co-spectra for each of the zones \citep{2018ApJS..236...13H}. The cospectrum is the real part of the cross-spectrum, and these were generated using 32~s segments with \texttt{Stingray} utilizing data from the two independent detectors (FPMA and FPMB); in this case, uncorrelated variability such as dead time effects, which can be significant for \nustar, is eliminated \citep{2018ApJS..236...13H,2019JOSS....4.1393H,2019ApJ...881...39H}. The cospectra were well described by a zero-centred Lorentzian (i.e., centroid frequency of $0$~Hz) to describe the broadband noise and a Lorentzian to describe the QPO.  In the case of zones 5 and 6, the narrow Lorentzian parameters were unconstrained, perhaps due to low statistics. We report the timing parameters from the cospectra in Table~\ref{tab:z-resolved-pds}. 

\begin{table}
  \centering
  \caption{Power density spectral properties of five HID zones (see Figure~\ref{fig:hid}). Observed QPOs, their fractional rms, and the total PDS rms in the frequency range of 0.05--100~Hz are provided. } 
  \label{tab:z-resolved-pds}
  \resizebox{\columnwidth}{!}{
\begin{tabular}{l|cc|cc}
\hline
Zone &  \multicolumn{2}{c|}{Observed QPO Frequency (Hz)} & \multicolumn{2}{c}{Observed QPO rms (\%)} \\
Number & \astr & \nustar & \astr & \nustar \\
\hline
Zone 1 &  $23.37^{+0.18}_{-0.17}$ & $21.2\pm0.2$ & $8.45^{+0.46}_{-0.56}$ & $6.8\pm0.2$ \\
Zone 2 &  $26.35^{+0.23}_{-0.24}$ & $27.9\pm0.3$ & $7.39^{+0.29}_{-0.30}$ & $5.5\pm0.3$ \\
Zone 3  & $38.14^{+1.12}_{-1.26}$ & $32.8\pm0.5$ & $3.27^{+1.14}_{-0.83}$ & $4.2\pm0.3$ \\
       &    $29.15^{+0.87}_{-0.76}$ & ---  & $4.96^{+0.87}_{-0.73}$ & --- \\
Zone 4 &  $36.42^{+0.87}_{-0.76}$ & $38.3\pm0.6$ & $3.73^{+0.38}_{-0.46}$ & $3.5\pm0.3$ \\
Zone 5 &  $43.23^{+2.15}_{-2.11}$ & --- & $3.37^{+1.35}_{-1.14}$ & --- \\\hline

  \end{tabular}
  }
\end{table}

\subsection{Spectral modeling}\label{ssec:spec}

To identify the emission components in \src, we investigated the spectra using the \astr\ and \ixpe\ observations,  as these were exactly simultaneous. 
The last three orbits of the SXT observations indicated an abnormal excess in count rate, which is not seen in the same bands in LAXPC or \ixpe\ light curves. Thus, we excluded those orbits from both LAXPC and SXT spectra.  We use \textsc{xspec} v12.14.0 to model the spectra, and for spectral and spectro-polarimetric modeling, we used the $\chi^2$ statistic.
Inferring from \citetalias{Bhargava2023ApJ...955..102B}, we tested if various combinations of spectral components, primarily the EM and the WM, are sufficient to describe the spectrum of the source. Noting a high absorption and typical modeling of harder components as a Comptonized emission, we initially modeled the spectra with \texttt{tbabs*nthcomp}. 
The absorption column was found to be consistent with 10$^{23}$~cm$^{-2}$ (\citealt{Iaria2006ChJAS...6a.257I,Church2006A&A...460..233C}, \citetalias{Bhargava2023ApJ...955..102B}) and we utilized the abundances from \citet{Wilms2000ApJ...542..914W} and cross-sections from \citet{Vern1996ApJ...465..487V} in the spectral modeling. For the \astr-SXT and \astr-LAXPC spectra, we observed that the $\chi^2$/degrees of freedom (dof) = 575/110 was unacceptable with clear residuals in the soft X-ray band hinting at the presence of a thermal component. For testing the EM/WM, we included a \texttt{diskbb}/\texttt{bbodyrad} to model the accretion disk/blackbody component, respectively. The resulting test statistic was lower (223.6/109 for WM and 513.9/109 for EM) but still unacceptable.
The spectral model \texttt{thcomp} \citep{thcomp2020MNRAS.492.5234Z} is a convolution version of \texttt{nthcomp} with improved implementation of relativistic effects and additional treatment of partial covering of the seed emission by the corona, and, therefore, for the spectral modeling, we replaced the \texttt{nthcomp} with the \texttt{thcomp} component. The \texttt{thcomp} model has also been used in spectro-polarimetric analysis of various weakly magnetized NS LMXBs \citep{Fabiani2024A&A...684A.137F,tarana2025A&A...698A.245T}.  In the case of \texttt{thcomp}, the key spectral parameters (i.e., the photon index ($\Gamma$) and electron temperature (kT$_e$)) are similar in description to the ones in \texttt{nthcomp}, with an additional covering fraction ($\mathfrak{f}$) parameter which accounts for the scattering of a fraction of seed photons. 
We observed a clear residual in the softer X-rays. As such, we included another thermal component\footnote{Seen in spectral modeling irrespective of the choice of the Comptonization model, \texttt{nthcomp} or \texttt{thcomp}.}. 
In the EM, the secondary thermal component was modeled as a blackbody, and in the WM, it was modeled as an accretion disk. On inclusion of the secondary component, the test statistic improved significantly and was observed to be similar for both WM and EM. 
For the EM, the description looks like \texttt{tbabs*(bbodyrad+thcomp$\otimes$diskbb)} and for the WM, it is \texttt{tbabs*(diskbb+thcomp$\otimes$bbodyrad)}. We present the parameters in Table~\ref{tab:spec_pars}. The confidence intervals were computed using Markov chain Monte Carlo (MCMC) sampling of the $\chi^2$ using 20 walkers for a total of 50000 steps (after discarding the initial 2000 steps as part of the burn-in phase).

\begin{figure*}
\centering
  \includegraphics[width=1.6\columnwidth]{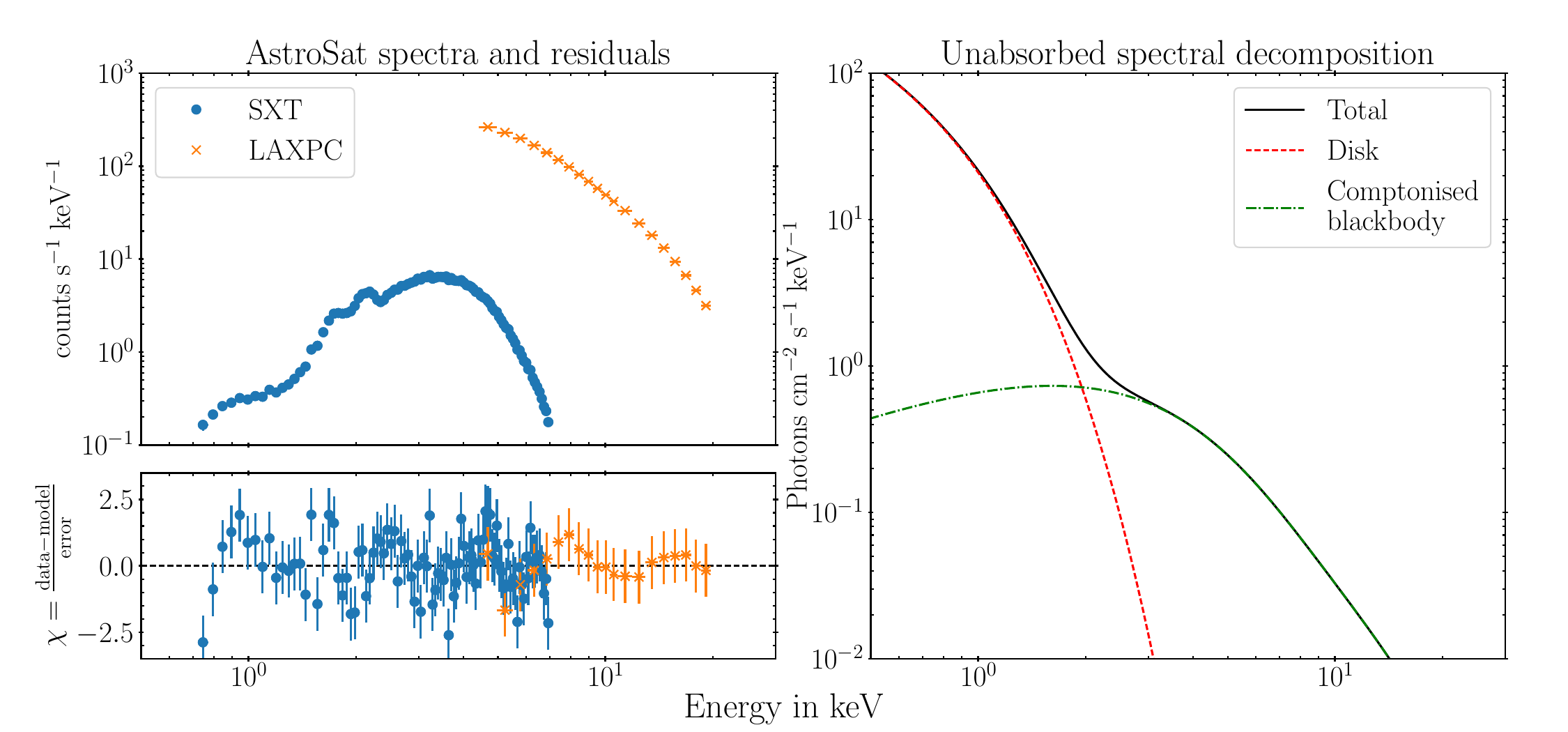}
  \caption{ \textit{Left panels}: \astr\ spectra (top) and residuals from the WM modeling (\texttt{tbabs*(diskbb+thcomp$\otimes$bbodyrad)}) of the \astr\ SXT and LAXPC spectra (bottom). \textit{Right panel}: Relative contributions of individual components for the WM are indicated as various lines, with the black solid line indicating the total emission. } \label{fig:spectra_astr}
\end{figure*}

\begin{table*}
  \centering
  \caption{Best-fit parameters from the spectral modeling of \astr\, \ixpe, and \nustar\ spectra individually. 
  The \astr\ spectra have been modeled with either WM: \texttt{tbabs*(diskbb + thcomp$\otimes$bbodyrad)} or EM: \texttt{tbabs*(bbodyrad + thcomp$\otimes$diskbb)}, while the \ixpe\ spectra have been modeled with only the WM. The \nustar\ spectra have been modeled with \texttt{tbabs*(diskbb + relxillNS + thcomp$\otimes$bbodyrad)} due to the presence of strong residuals at $\sim6.4$~keV}.    
  \label{tab:spec_pars}

  \resizebox{1.7\columnwidth}{!}{
  \begin{tabular}{|l|c|c|c|c|c|c|}
    \hline
    \multicolumn{3}{|c|}{} & \multicolumn{2}{c|}{\astr} & {\ixpe} & \nustar \\ \hline
    Component & Parameter & Unit & WM & EM & WM & WM \\ \hline
    
    \texttt{tbabs} & $n_{\rm H}$ & $10^{22}$ cm$^{-2}$ & $9.8_{-0.4}^{+0.4}$ & $10.2_{-0.4}^{+0.4}$ & $8.3_{-0.4}^{+0.4}$ &   $9.8^\star$ \\ \hline
        \texttt{diskbb} & kT$_{\rm disk}$ & keV        & $0.252_{-0.011}^{+0.007}$ & $0.227_{-0.007}^{+0.006}$ & $0.252^\star$ &  $0.79_{-0.01}^{+0.01}$\\
         & Norm$^\ast$ &                               & $8_{-2}^{+4}\times10^{5}$ & $2.2_{-0.4}^{+0.5}\times10^{6}$ & $6_{-4}^{+8}\times10^{4}$ &   $1.9_{-0.1}^{+0.1}\times10^3$ \\ \hline
    \texttt{bbodyrad} & kT$_{\rm bb}$ & keV            & $1.02_{-0.04}^{+0.02}$ & $1.09_{-0.02}^{+0.03}$ & $0.65_{-0.02}^{+0.04}$ &   $1.33_{-0.01}^{+0.01}$\\
     &Norm$^\dagger$ &                                 & $1150_{-90}^{+230}$ & $580_{-80}^{+70}$ & $3.6_{-0.5}^{+0.6}\times10^3$  & $ 325_{-10}^{+10}$\\ \hline
    \texttt{thcomp} & $\Gamma$ &                      & $2.1_{-0.3}^{+0.4}$ & $1.6_{-0.6}^{+1.0}$ & $1.88_{-0.11}^{+0.05}$ &   $2.25_{-0.07}^{+0.15}$\\
      & kT$_{e}$ & keV                & $3.5_{-0.4}^{+0.9}$ & $3.1_{-0.1}^{+0.2}$ & $4_{-1}^{+2}$  & $3.05_{-0.04}^{+0.11}$ \\
     & $\mathfrak{f}$ &               & $0.45_{-0.11}^{+0.17}$ & $0.015_{-0.002}^{+0.005}$ & $0.96_{-0.1}^{+0.03}$ &   $0.79_{-0.05}^{+0.07}$\\ \hline
    \texttt{relxillNS} & A$_{\rm Fe}$ &  & ---&---&---&  $0.6_{-0.1}^{+0.3}$ \\
     & Log $\xi$ & &---&---&---&  $3.00_{-0.05}^{+0.05}$ \\
     & Norm$^{\star\star}$ & &---&---&---&   $4.2_{-0.6}^{+0.9}\times10^{-3}$ \\\hline
    Gain parameters & SXT Offset & keV & 0.015 & 0.015 & --- & --- \\
     & DU1 Slope & & ---& ---& $1.009$ & ---\\
     & DU1 Offset & keV &--- &--- & $0.003$ &--- \\
     & DU2 Slope & &--- &--- & $0.996$ &--- \\
     & DU2 Offset & keV &--- &--- & $0.071$ &--- \\
     & DU3 Slope & &--- &--- & $1.001$ &--- \\
     & DU3 Offset & keV &--- &--- & $0.043$ &--- \\ \hline
    Test statistic & $\chi^2$/dof & & 116/108&  117/108 & 435/432 &  1294/1108 \\ \hline
    
  \end{tabular}

  }
  \begin{flushleft}
    Notes: \\
    The 90\% confidence intervals ($\Delta\chi^2=2.706$) for each parameter are noted above. \\
    $\ast$: The disk normalization is defined as Norm $=(R_{eff}/D_{10})^2*\cos \theta $, where $R_{eff}$ is the apparent inner radius of the accretion disk in km, $D_{10}$ is the distance to the source in units of 10~kpc, and $\theta$ is the inclination angle of the disk in degrees.  \\
    $\dagger $: The blackbody normalization is defined as Norm $=(R/D_{10})^2$, where $R$ is the radius of the emission region in km.\\
    $\star$: The parameter is kept fixed \\
    $\star\star$: The normalization of \texttt{relxillNS} component is the defined in \citet{relxill2016A&A...590A..76D} and \cite{relxillns2022ApJ...926...13G}. \\
  \end{flushleft}

\end{table*}

Similarly, if we model the \ixpe\ spectra starting with individual components, we observe that the emission can be adequately modeled with just absorbed Comptonized emission (i.e. \texttt{tbabs*nthcomp}) for a test-statistic of $\chi^2$/dof = 440.6/438. For the best fit, we leave the gain parameters of all DUs free \citep[e.g.][]{Fabiani2024A&A...684A.137F, Steiner2024ApJ...969L..30S}.
For a comparison with the spectral modeling of broadband \astr\ spectra and subsequent spectro-polarimetric modeling, we finally modeled the \ixpe\ spectra with the WM, i.e., \texttt{tbabs*(diskbb+thcomp$\otimes$bbodyrad)}.  
We also conducted some preliminary cross-instrument comparisons between \astr\ and \ixpe\ as this is the first time that only these instruments are simultaneously observing a target. The details of the tests are reported in appendix~\ref{appsec:crosscal}. On modeling the \ixpe\ spectra with WM, we find the spectral parameters differ from those observed with \astr. This could be due to 1) spectral calibration mismatch or 2) spectral evolution. 
The \astr\ observation samples a part of the HB uniformly, while the \ixpe\ observation almost completely covers the HB but non-uniformly. Since there is a strong dependence of spectral parameters of source on the Z-track position (\citealt{Seifina2013ApJ...766...63S}, \citetalias{Bhargava2023ApJ...955..102B}), it is likely that \ixpe\ spectral modeling will yield significantly different parameters as compared to \astr\ even after correction for calibration mismatch. Since for polarimetric analysis, we are combining the complete observation, we consider the same for spectral modeling of \ixpe\ data. 
Thus we have modeled the \ixpe\ spectra mostly independently. However, we fix the kT$_{\rm disk}$ parameter as to the \astr\ model as \ixpe\ doesn't cover the energy range to constrain it. 
We have noted the spectral parameters in Table~\ref{tab:spec_pars}. 
For spectro-polarimetric modeling (section~\ref{ssec:specpol}), we utilize the spectral parameters from the modeling of the \ixpe\ spectra with the WM.

In addition to the \astr\ and \ixpe\ observations, our campaign also included observations from \nicer, \hxmt, and \nustar. However, these observations were not simultaneous with our \ixpe\ observation. As such, we haven't included detailed spectral investigations using these observations. The timing properties of \nustar\ observations strongly suggest that these also occurred during the HB. Therefore, we attempted spectral modeling on the \nustar\ data to investigate if its sensitivity allows for constraints on the reflection features which have been historically observed in \src\ \citep{dai2009ApJ...693L...1D, 2010ApJ...720..205C,miller2016ApJ...822L..18M, monaca2024A&A...691A.253L}.

The \nustar\ observations traced the HB multiple times. To compare the spectral modeling with the \astr\ and \ixpe\ data, we modeled the \nustar\ observation 91002313008, which traced the largest extent of the track (see Figure~\ref{fig:hid}). Additionally, the multiple visits during different parts of the HB allowed the evolution of the reflection features to be traced as the source moved along the HB. Therefore, we also investigated the spectra from resolved individual HID zones (as divided in Figure~\ref{fig:hid}), where we combined the spectra from different observations that lie in the same zone, providing better statistics. 
We employed the same recipe as described above, determining the minimum number of components needed to model the spectra. Due to some minor discrepancy between the FPMA and FPMB at energies $<$4~keV, we modeled the \nustar\ spectra in only the 4--79~keV energy range.
We found that the continuum emission in the \nustar\ spectra is well modeled with the WM (i.e., with \texttt{tbabs*(diskbb+thcomp$\otimes$bbodyrad)}) but shows strong asymmetric residuals at $\sim6.4{\rm\,keV}$. 
The asymmetric feature is observed in the six zone-resolved spectra and the spectra from observation 91002313008, and we interpret it as the reflection component. 
To model the reflection component, we performed preliminary fits with continuum as described earlier and included a simple Gaussian, a \texttt{diskline} model, a \texttt{relline} model, and  \texttt{relxillNS} \citep{1989MNRAS.238..729F,nthzdz1996MNRAS.283..193Z,zycki1999MNRAS.309..561Z,2010MNRAS.409.1534D,relxillns2022ApJ...926...13G}. 
We found that the \texttt{relxillNS} model best describes the reflection features, as determined by the $\Delta\chi^2$ between the inclusion and exclusion of the model \citep{2010MNRAS.409.1534D}. 
We fix the emissivity index to 3, the outer disk radius to 400~$R_{\rm g}$, the disk density to $10^{19}{\rm\,cm^{-3}}$, spin parameter to 0.0, \citep{monaca2024A&A...691A.253L}, and the inclination to 35$^\circ$  (as reported by \citealt{miller2016ApJ...822L..18M} from high-resolution \textit{Chandra}/HETGS X-ray observations), leaving only iron abundance (A$_{\rm Fe}$), ionization parameter ($\xi$) and the normalization free. We also tied the relevant parameters of the reflection component to the thermal emission to link the seed photon source and set the reflection parameter to -1 to only model the reflected emission.
We report the spectral parameters for the modeling of the \nustar\ observation 91002313008 in Table~\ref{tab:spec_pars}. 

To characterize the flux from the emission line across the six zone-resolved \nustar\ spectra, we also modeled the spectra with the same continuum model and a relativistic line model \texttt{relline}. 
We find that the rest-frame centroid energy of the feature did not significantly shift as a function of the HID zone. We found the line energy varies between 6.36~keV and 6.58~keV, consistent within $2\sigma$. However, the line normalization decreased (though not monotonic) from $(5.9\pm0.6)\times10^{-3}{\rm\,photons\,cm^{-2}\,s^{-1}}$ in zone 1 to $(1.8_{-1.2}^{+1.3})\times10^{-3}{\rm\,photons\,cm^{-2}\,s^{-1}}$ in zone 6. 
Since the \nustar\ observations were not exactly simultaneous with the \ixpe\ observations, we did not perform a joint spectro-polarimetric analysis that would have provided additional insight into the polarization properties of the reflection component  (however, see \citealt{monaca2024A&A...691A.253L} for a reflection-based polarimetric study of the source).

\subsection{Spectro-polarimetric investigation}\label{ssec:specpol}

The spectral analysis revealed that multiple components contribute to the total emission in the 2--8~keV band, while the model-independent polarimetric analysis suggested these components could contribute differently to the total polarization, particularly in the 2--2.5~keV band \citep[also seen for other Z-sources;][]{Farinelli2023MNRAS.519.3681F, Fabiani2024A&A...684A.137F}. Motivated by this, we conducted a spectro-polarimetric analysis using the \ixpe\ observations.

We jointly modeled the Stokes I, Q, and U spectra for all the \ixpe\ detector units to calculate the spectro-polarimetric results. We fixed the spectral parameters to the values from the modeling of the Stokes I spectra and included multiplicative component(s) to infer the polarimetric properties from the Stokes Q \& U spectra.  Firstly,  we checked if the observation was consistent with a single polarization value for all components by fitting the spectra to \texttt{tbabs*polconst*(diskbb + thcomp$\otimes$bbodyrad)}  (with the spectral parameters fixed at the values mentioned in Table~\ref{tab:spec_pars}). 
We found that assuming a constant polarization resulted in an acceptable fit ($\chi^2$/dof=1243/1336) and a PD/PA of $3.6\pm0.3$\%/$36\pm2^\circ$ (consistent with the model-independent analysis in section~\ref{ssec:pcube}). Since the model-independent analysis indicated a possible change in the PA as a function of energy, we replaced the constant polarization with linear polarization (\texttt{pollin}), keeping the slope of the PD fixed at 0 and the slope of the PA free. It resulted in a slightly lower $\chi^2$/dof = 1236/1335. We report the best-fit spectro-polarimetric parameters in Table~\ref{tab:pol_spec}.
To determine which model is statistically better, we computed the Akaike Information Criterion (AIC, \citealt{Akaike1974ITAC...19..716A})  and Bayesian Information Criterion (BIC) for both models\footnote{The key difference between these criteria is the penalization of likelihood (as computed from the $\chi^2$ statistic) with the number of parameters. If both criteria agree/disagree, then it is valid to reject/accept the null hypothesis. But if they result in conflicting observations, then it is preferable to assume that the simpler model is better. }.  
 Both AIC and BIC prefer \texttt{pollin} over \texttt{polconst} with strong evidence ($\Delta$AIC$\approx$$\Delta$BIC=6.9), but do not support it overwhelmingly (i.e. $\Delta$IC$\gtrsim$10). Thus we prefer the simpler model of the two, i.e. \texttt{polconst}.

\begin{table}
    \centering
        \caption{Spectro-polarimetric properties of \src. The 1$\sigma$ confidence intervals are reported. Spectral parameters are fixed for the WM model for \ixpe\ as reported in Table~\ref{tab:spec_pars}. }    \label{tab:pol_spec}
    \begin{tabular}{l|r}
         \multicolumn{2}{c}{Constant polarization
         } \\ \hline         
         Parameter & Value \\ \hline
         PD & $3.69\pm0.3\%$\\
         PA &  $36\pm2^\circ$\\ \hline 
         \multicolumn{2}{c}{Linear dependence of PA on energy
         } \\ \hline
         Parameter & Value \\ \hline
         PD at 1 keV  & $3.7\pm0.3\%$\\
         PD slope&  0\%/keV(fixed)\\ 
         PA at 1 keV  & $20\pm6^{\circ}$\\
         PA slope&  $5\pm2^\circ$/keV\\\hline 
         \multicolumn{2}{c}{Polarization from individual components
         } \\ \hline
         Parameter & Value \\ \hline
         Disk PD & $1\%$ (fixed)\\
         Disk PA &  $-27_{-35}^{\circ +85}$\\ 
         Comptonized PD & $3.6\pm0.3\%$\\
         Comptonized PA &  $36\pm2^\circ$\\ \hline
    \end{tabular}
\end{table}

Other Cyg-like Z-sources have hinted at a possible difference in the PA of additive components \citep{Farinelli2023MNRAS.519.3681F, Fabiani2024A&A...684A.137F}. Therefore, we tested this by applying \texttt{polconst} individually to \texttt{diskbb} and    \texttt{thcomp$\otimes$bbodyrad}. Since the disk emission is majorly obscured by the intervening medium, we fix the PD of the disk component to a value expected from an accretion disk inclined at $35^\circ$, i.e., 1\% \citep[e.g.][]{1960ratr.book.....C, Li2009ApJ...691..847L}. 
We found that with this assumption, the fit is also slightly improved as compared to a single polarization component ($\chi^2$/dof=1243/1335). The PA of the accretion disk component prefers a value of $-27_{-35}^{\circ+85}$ ($1\sigma$ confidence) while the Comptonized emission has PD: $3.6\pm0.3\%$ and PA: $36\pm2^\circ$ ($1\sigma$ confidence). We note that the PAs of both components are consistent within $1\sigma$ confidence interval. 
To compare the spectro-polarimetric fit with individual polarization components to the single polarization component, we computed the AIC and BIC for both models. 
We found that both criteria suggest that either model is equally good, and since a single polarization component has fewer parameters, we prefer that assumption.
We have also summarized the results from all cases tested in this section in Table~\ref{tab:pol_spec}.

\subsection{Results from Radio campaign}

Our \ixpe\ observations of \src\ were accompanied by a supporting quasi-simultaneous radio campaign with the \gmrt\ and \atca. See section~\ref{appsec:data_red} for the complete details of the observations, data reduction, and analysis. The \gmrt\ observations were carried out on two epochs, on 2024-03-20/21 and 2024-03-27/28 at central frequencies of 750~MHz and 1.26~GHz on both dates (see Table~\ref{tab:obslog}). On both dates, in both frequency bands, no radio counterpart to \src\ was detected by the \gmrt. 

\atca\ observations occurred on 2024-03-29 and 2024-04-01, where data were taken at 5.5~GHz and 9~GHz. A bright radio source was detected at both frequencies on both dates with flux densities of $720 \pm 60$~$\mu$Jy at 5.5~GHz and $385 \pm 50$~$\mu$Jy at 9~GHz on 2024-03-29, and $720 \pm 30$~$\mu$Jy at 5.5~GHz and $595 \pm 30$~$\mu$Jy at 9~GHz on 2024-04-01. These flux densities are brighter than previously reported limits taken at similar frequencies \citep{2000MNRAS.318..599B}. No linear polarization (LP) or circular polarization (CP) was detected, with 3$\sigma$ upper limits of $<$6\% on the LP and $<$4\% on the CP (centred at 7.25 GHz). These constraints are consistent with expectations for the radio emission from an accreting X-ray binary jet, where values of a few percent are typically reported for LP, and CP measurements are rare or low \citep[e.g.,][]{2006csxs.book..381F}.

\section{Discussion}

During our multiwavelength campaign, \src\ remained primarily in the HB. The HIDs from various X-ray instruments, the timing analysis of \astr\ and \nustar\ observations (which showed QPOs at frequencies typical of HB oscillations), and the spectral properties of the source provide ample evidence for the state of the source. The HB oscillation evolved across the HB in a typical manner, i.e., increased as the source moved along the HB from the lowest flux in the HB to the apex, from $\sim$21.2--23.4~Hz to $\sim$43.2 Hz while the fractional rms amplitude of the QPO decreased (see figure~\ref{fig:dps}). Similar timing evolution has previously been reported using various fast-timing instruments (\citealt{Jonker2000ApJ...537..374J, Jonker2002MNRAS.333..665J}, \citetalias{Bhargava2023ApJ...955..102B}, \citealt{Pahari2024MNRAS.528.4125P}) and similar HBO frequencies were found. The hard apex is expected to show a frequency of $\sim45$~Hz \citep{Jonker2000ApJ...537..374J} and all QPOs we observe are lower in frequency, which indicates during \astr, \nustar, and by extension the \ixpe\ observations, the source was in the HB.  \src\ staying within the HB for a duration of $\sim$ 13 days is quite unusual, where it typically evolves through its complete HID over a time scale of a few days (\citealt{Jonker2000ApJ...537..374J} and recently, \citealt{Bhargava2024arXiv241100350B}). Although it is possible that during the intervals where we did not have pointed X-ray observations, it may have transitioned to the NB/FB and returned, but the \textit{MAXI}/GSC monitoring of the source is unable to discern such an evolution. 

The broadband spectral investigation with \astr\ presents three emission components in the spectra with a preference to a modified WM, i.e., with a soft multicolor disk, a hotter blackbody-like emission, and a Comptonized emission that up-scatters seed photons from the blackbody component. 
The disk emission typically contributes at $\lesssim3$~keV, while the blackbody and Comptonized emission have roughly similar contributions (as suggested by the value of the covering fraction in the spectral modeling) in the 3--8~keV band. Our analysis suggests that if the source of the seed photons for the Comptonization is an accretion disk (as expected in EM), only a small fraction of the photons are extensively up-scattered to produce the observed non-thermal emission. This implies that the corona does not cover the disk. On the other hand, if the source of the seed photons is the blackbody-like emission (perhaps from the NS surface or the BL/SL), the covering fraction is much more reasonable, with roughly equal contribution of the blackbody and thermal emission, which is also consistent with the literature \citep[e.g.,][]{Church2006A&A...460..233C}. In this paradigm, the corona can be imagined to be a slab-like structure between the disk and the NS and intercepts a fraction of photons from the NS surface. 

The \ixpe\ spectra in the 2--8~keV energy range only require a single Comptonized component for an adequate description (as noted by an acceptable test statistic), but for a better comparison with the broadband spectra, we model the spectro-polarimetric components with a WM description with key parameters fixed at those found in modeling of the \astr\ spectra. The reflection component observed in the \nustar\ spectra is consistent with the previous observations of the source \citep[e.g.][]{miller2016ApJ...822L..18M, monaca2024A&A...691A.253L}, suggesting that the source was observed in its typical HB state.

\subsection{Discovery \& Nature of X-ray Polarization in GX 340+0}

The first \ixpe\ observation of \src\ revealed a significant X-ray polarization in the 2--8~keV energy band. The PD of the source in the HB was measured to be $4.02 \pm 0.35$\%. A similar PD has been reported from XTE J1701$-$462 \citep{Cocchi2023A&A...674L..10C} and GX~5$-$1 \citep{Fabiani2024A&A...684A.137F}. The PA of the polarized emission across the full 2--8\,keV energy band was measured to be $37.6 \pm 2.5\degr$.  Similar values have also been reported in independent analyses \citep{monaca2024A&A...691A.253L, Gnarini2025A&A...699A.230G}, from the same \ixpe\ observation. 

The energy-resolved polarization study indicates that at the lowest energies (i.e., 2--2.5~keV), \src\ shows a different (albeit marginal) PA as compared to the PA of the rest of the emission (see Figure~\ref{fig:pcube}). 
Historically, at energies $\gtrsim 3$~keV, two spectral components have been found to dominate the emission \citep{Jonker2002MNRAS.333..665J, lin2009ApJ...696.1257L, lin2012ApJ...756...34L}. The model-independent observation of polarization indicates that in the higher bands ($\gtrsim 2.5$~keV) all of the emission aligns in PA. These components have been modeled as a combination of accretion disk+Comptonized emission (for Cyg X-2; \citealt{Done2002MNRAS.331..453D}), blackbody+Comptonized emission (\citealt{Church2006A&A...460..233C}, \citetalias{Bhargava2023ApJ...955..102B}, and present work, Figure~\ref{fig:spectra_astr}), or a combination of two Comptonized emissions of different electron temperatures \citep{Seifina2013ApJ...766...63S}. Since we do not observe a significant variation of the PA across the higher 2.5--8~keV energies, we can infer that either both components have an aligned emission or one component has a stronger/only contribution to the polarized emission, dictating the observed polarization properties. The latter scenario matches the model incorporated in our spectral analysis, i.e., Comptonized blackbody emission, in which the polarization is dominated by up-scattered emission, while the thermal component is expected to be unpolarized or weakly polarized \citep[e.g.,][and references therein]{Farinelli2024A&A...684A..62F, Bobrikova2025A&A...696A.181B}. At energies in the 2--2.5~keV band, the net PA of the emission changes marginally ($<3\sigma$ confidence).  Simulations by \citet{gnarini2022MNRAS.514.2561G} for a slab-like corona predict a change in the observed PA similar to what is observed in \src\ (i.e., lower PA in softer X-rays) for various inclination angles.

For our spectro-polarimetric analysis, we have assumed that the emission comprises an accretion disk and a Comptonized blackbody. 
The joint Stokes-parameter spectra can be sufficiently explained by a single polarized component. This is also consistent with the observed spectral decomposition in 2--8~keV as the emission is dominated by the Comptonized blackbody. The disk emission has a small contribution in 2--2.5~keV, and we tested if its polarization properties can be reliably disentangled.
Assuming constant polarization from individual components and fixing the PD from the accretion disk inclined at $35\degr$ \citep{miller2016ApJ...822L..18M} to the expected value of 1\% \citep{Li2009ApJ...691..847L}, we estimate that the disk component has a PA consistent with the Comptonized emission due to the large uncertainty on the PA. The high uncertainty in the estimation can be attributed to a few factors: a) low modulation response of the \ixpe\ detectors \citep{dimarco2022AJ....164..103D}, and  b) low disk contribution in 2--2.5~keV and c) statistics of the spectral modeling (i.e. \ixpe\ spectra can be modeled well with a simple absorbed Comptonized emission without requiring a disk component, section~\ref{ssec:spec}). 
However, other Cyg-like Z sources have shown differences in the PA of disk and Comptonized emission \citep{Cocchi2023A&A...674L..10C, Fabiani2024A&A...684A.137F}, which could indicate that it might be an intrinsic property of the HB of the Cyg-like Z-sources. The scattering from the disk wind and intrinsic polarization from the accretion disk is expected to be perpendicular to the polarization of the Comptonized emission (\citealt{Sunyaev1985A&A...143..374S}, and with hints of 90$\degr$ offset in Cyg X-2, \citealt{Farinelli2023MNRAS.519.3681F}) but a lower observed discrepancy could indicate that strong GR effects may play a role as well \citep[e.g.][]{schnittman2009ApJ...701.1175S}.  
Given a lack of robust disentangling of the disk PA and Comptonized emission PA, such effects cannot be interpreted for this source. Any position angle constraints from radio wavelengths (e.g., detection of a jet direction or estimate of the radio polarization angle) should provide additional information about the relative geometrical orientation of the accretion disk and the Comptonising medium.  
The degree of polarization expected from disk+corona systems as simulated by \citet{gnarini2022MNRAS.514.2561G} suggests that an inclination angle of $\gtrsim60\degr$  is required to produce the $\sim4\%$ PD, higher than the inclination estimated from reflection studies \citep[$\sim35\degr$;][]{miller2016ApJ...822L..18M} which was also noted by \citep{monaca2024A&A...691A.253L}. This discrepancy could be explained by a potential misalignment between the inner regions of the accretion disk (that is covered by the slab-like corona) and the accretion disk probed by the reflection component.

\subsection{Radio emission from GX~340+0}

The non-detection of \src\ in the \gmrt\ observations but detection by the \atca\ observations suggests a break in the radio spectrum, lying between the 1.26~GHz \gmrt\ band and the 5.5~GHz \atca\ band. Such a break can be a result of either synchrotron self-absorption or free-free absorption \citep[e.g.,][]{1974ApJ...194..715G,1976ApJ...207...88S,2004ApJ...600..368M,2021PASA...38...45C}. 
However, due to the non-detection in the lower frequency bands ($<$2 GHz), we are unable to determine which mechanism is responsible for the spectral turnover \citep[e.g.,][]{1986rpa..book.....R}. We also note that the radio source would be expected to vary as the accretion properties evolve on relatively short timescales, as seen in the evolution of radio flux at 9~GHz across two \atca\ observations. 
As the \gmrt\ and \atca\ observations are not exactly simultaneous, it is, therefore, possible that the source state has changed between the observations. 
However, during both \gmrt\ observations, we have simultaneous X-ray coverage (\nicer\ during the first and \hxmt\ during the second, see Figure~\ref{fig:full_lc}), and the X-ray HID in Figure~\ref{fig:hid} clearly indicates that during both these epochs the source was on the HB. 
The second \atca\ observation is strictly simultaneous with \nustar, also residing on the HB as well.
Previous investigations in radio wavelengths have suggested that the radio emission can also be highly variable (and correlated with the X-ray variability). Comparing our results with historical radio studies showing a non-detection with upper-limits of 200~$\mu$Jy \citep{2000MNRAS.318..599B} and 1.0--1.5~$\mu$Jy detections during the NB \citealt{oosterbroek1994A&A...281..803O} clearly indicate that the radio emission is indeed variable between states. Hence, simultaneous observations of \src\ are paramount to determine the behavior of the source.

\section{Conclusions}

Here, we summarize the key results from our X-ray and radio investigation of \src\ primarily with \ixpe\ to investigate the polarization properties. 

\begin{itemize}
    \item The source stayed in the horizontal branch during our multiwavelength observations, which spanned roughly 13 days.

    \item For the first time, we detect X-ray polarization from \src, with a 2--8~keV PD of $4.02 \pm 0.35$\%   and PA = $37.6 \pm 2.5\degr$.
We also detect significant polarization in various energy sub-bands. The 2--2.5~keV energy sub-band polarization hints at a different PA (at a $2.8\sigma$ confidence) as compared to the PA of other energy bands (which all have a consistent PA).

    \item The broadband spectra required 3 components (accretion disk, blackbody, and Comptonized emission) to adequately describe the X-ray spectra. The accretion disk is found to have a low temperature ($\sim0.2$~keV). The blackbody component is observed to have a temperature of $\sim1.1$~keV, and the Comptonizing medium is observed to scatter $\sim0.65$ of the seed photons from the blackbody component. Additionally, the \nustar\ spectra show an asymmetric iron line feature consistent with previous reports of reflection features, which varies as the source moves along the HB.  

    \item Due to almost all energy bands depicting a similar PA, a constant polarization (i.e., no dependence of PD/PA on energy) can fit the spectro-polarimetric data. The slight deviation in the PA in the lowest energy band can be explained by a PA varying linearly with the energy, but the data is not sufficient to rule out constant polarization.

\item At radio wavelengths, the source was not detected between 0.7--1.5 GHz (3$\sigma$ upper limit of 0.9~mJy) but was significantly detected at higher frequencies (5.5--9 GHz). This is indicative of a spectral break somewhere in the 1.5--5.5 GHz range. Additionally, linear polarization was not detected at radio wavelengths, with a  3-sigma upper-limit of $<$6\%, and circular polarization is $<$4\%. 
\end{itemize}

\begin{acknowledgements}
The authors would like to thank the anonymous referees for their valuable comments, which have significantly improved the article. The Imaging X-ray Polarimetry Explorer (IXPE) is a joint US and Italian mission.  The US contribution is supported by the National Aeronautics and Space Administration (NASA) and led and managed by its Marshall Space Flight Center (MSFC), with industry partner Ball Aerospace (now, BAE Systems).  The Italian contribution is supported by the Italian Space Agency (Agenzia Spaziale Italiana, ASI) through contract ASI-OHBI-2022-13-I.0, agreements ASI-INAF-2022-19-HH.0 and ASI-INFN-2017.13-H0, and its Space Science Data Center (SSDC) with agreements ASI-INAF-2022-14-HH.0 and ASI-INFN 2021-43-HH.0, and by the Istituto Nazionale di Astrofisica (INAF) and the Istituto Nazionale di Fisica Nucleare (INFN) in Italy.  This research used data products provided by the IXPE Team (MSFC, SSDC, INAF, and INFN) and distributed with additional software tools by the High-Energy Astrophysics Science Archive Research Center (HEASARC) at NASA Goddard Space Flight Center (GSFC).
This work makes use of data from the AstroSat mission of the Indian Space Research Organisation (ISRO), archived at the Indian Space Science Data Centre (ISSDC). The article has used data from the SXT and the LAXPC developed at TIFR, Mumbai, and the AstroSat POCs at TIFR are thanked for verifying and releasing the data via the ISSDC data archive and providing the necessary software tools. This work has made use of data from the \nustar\ mission, a project led by the California Institute of Technology, managed by the Jet Propulsion Laboratory, and funded by the National Aeronautics and Space Administration. We thank the \nustar\ Operations, Software, and Calibration teams for support with the execution and analysis of these observations. This research has made use of the \nustar\ Data Analysis Software (NUSTARDAS) jointly developed by the ASI Science Data Center (ASDC, Italy) and the California Institute of Technology (USA). This work was supported by NASA through the \nicer\ mission and the Astrophysics Explorers Program. 
This work has made use of the data from the \hxmt\ mission, a project funded by the China National Space Administration (CNSA) and the Chinese Academy of Sciences (CAS).
This research has also made use of data and/or software provided by the High Energy Astrophysics Science Archive Research Center (HEASARC), which is a service of the Astrophysics Science Division at NASA/GSFC and the High Energy Astrophysics Division of the Smithsonian Astrophysical Observatory. We thank the staff of the GMRT who made these observations possible. GMRT is run by the National Centre for Radio Astrophysics of the Tata Institute of Fundamental Research. The Australia Telescope Compact Array is part of the Australia Telescope National Facility (\url{https://ror.org/05qajvd42}), which is funded by the Australian Government for operation as a National Facility managed by CSIRO. We acknowledge the Gomeroi people as the Traditional Owners of the ATCA observatory site. We thank Jamie Stevens and the ATCA staff for making the observations possible. YB and AB would like to especially thank Dr Shriharsh Tendulkar for valuable discussion on radio observations and insights into the emission from the neutron stars. YB would like to thank Dr. Andrzej Zdziarski for the useful discussion on the choice of Comptonizing models. YB would like to thank Dr. Felix F\"urst for the useful discussion on cross calibration of instruments. 
LZ acknowledges support from the National Natural Science Foundation of China (NSFC) under grant 12203052.  JH acknowledges support for this work from the IXPE Guest Investigator program under NASA grant 80NSSC24K1746.

\end{acknowledgements}

%

\bibliography{ref}
\bibliographystyle{aa}







   
  



\begin{appendix}
    \onecolumn
\section{Testing the spectral cross calibration of AstroSat and IXPE}\label{appsec:crosscal}
The \astr\ and \ixpe\ observations of \src\ have intervals of simultaneous overlap, and thus we could test the spectral cross-calibration between the instruments for the first time. For initial tests, we model solely the \ixpe\ observations with a simple model (\texttt{constant*tbabs*nthcomp}) to identify calibration differences between the individual detector units. The  residuals were observed to be quite strong and we mitigated these by leaving the gain parameter free, which allowed for an acceptable fit for all the detector units. 

During the simultaneous interval between \astr\ and \ixpe, the source count rate flucutated over time (Figure~\ref{fig:full_lc}), suggesting a movement of the source on the Z-track. Ideally, we should compare only the spectra across similar positions on the track (as different positions on the track have significantly different spectral parameters, e.g. \citetalias{Bhargava2023ApJ...955..102B}). The lightcurve of the observations suggested a strong overlap and near identical evolution in both instruments, while the  HID indicated that across the simultaneous interval, the source didn't do a significant back and forth tracing. To simplify the cross-calibration tests, we extracted spectra corresponding to the satellite orbits for both cases.  For each such interval, we extracted the  spectra from all the instruments and modeled them jointly. The spectral parameters are noted in Table~\ref{tab:crosscal}.  We found that for simultaneous intervals, we were able to obtain acceptable a fit statistic for joint modeling with the WM. 
The parameter values are similar to the ones reported to individual \astr\ and \ixpe\ modeling, which suggests that, overall, the model decomposition explains the observed spectra, while the parameter values are biased by the instruments' energy coverage, calibration, degrees of freedoms contribution in the joint modeling. The we can safely use the modeled parameters in Table~\ref{tab:spec_pars} for the spectro-polarimetric modeling as they represent the spectrum as observed by \ixpe. 

\begin{table}[ht!]
  \centering
  \caption{}\label{tab:crosscal}
  \begin{tabular}{|l|c|c|c|c|c|c|}
  \hline
  \multicolumn{3}{|c|}{} & \multicolumn{4}{c|}{Interval number}    \\ \hline
  Component & Parameter& Unit & 1&2&3&4\\ \hline
  \texttt{tbabs} & $n_{\rm H}$ & $10^{22}$ cm$^{-2}$  & $8.4_{-0.3}^{+0.3}$    & $8.4_{-0.4}^{+0.2}$    & $7.6_{-0.3}^{+0.2}$     & $8.4_{-0.3}^{+0.3}$\\ \hline
  \texttt{diskbb} & kT$_{\rm disk}$ & keV             & $0.19_{-0.01}^{+0.01}$ & $0.19_{-0.01}^{+0.01}$ & $0.18_{-0.01}^{+0.01}$  & $0.17_{-0.01}^{+0.01}$\\
      & Norm$^\ast$ &   $\times10^6$                  & $4_{-1}^{+2}$          & $3.2_{-0.9}^{+2.2}$    & $2.2_{-0.9}^{+1.4}$     & $9_{-3}^{+4}$\\ \hline
  \texttt{bbodyrad} & kT$_{\rm bb}$ & keV             & $1.14_{-0.02}^{+0.03}$ & $1.11_{-0.02}^{+0.03}$ & $1.17_{-0.02}^{+0.02}$  & $1.10_{-0.04}^{+0.03}$\\
   &Norm$^\dagger$ &                                  & $640_{-40}^{+60}$      & $670_{-70}^{+40}$      & $500_{-40}^{+40}$       & $660_{-70}^{+90}$\\ \hline
  \texttt{thcomp} & $\Gamma$ &                        & $2.6_{-0.1}^{+0.2}$    & $2.8_{-0.2}^{+0.2}$    & $2.1_{-0.2}^{+0.1}$     & $2.6_{-0.1}^{+0.1}$\\
   & kT$_{e}$ & keV                                   & $7.6_{-1.6}^{+2.4}$    & $10_{-3}^{+6}$         & $4.5_{-0.6}^{+0.4}$     & $6.5_{-1.1}^{+1.5}$\\
   & $\mathfrak{f}$ &                                 & $0.62_{-0.08}^{+0.08}$ & $0.77_{-0.13}^{+0.10}$ & $0.44_{-0.08}^{+0.06}$  & $0.65_{-0.05}^{+0.13}$\\ \hline
   Gain parameters & SXT Offset & keV                 & 0.06                   & 0.06                   & 0.05                    & 0.05\\
    & DU1 Slope &                                    & 0.98                    & 0.99                   & 0.99                    &1.00\\
    & DU1 Offset & keV                                & 0.01                   & -0.03                  & -0.04                   &-0.06\\
    & DU2 Slope &                                     & 0.97                   & 0.97                   & 0.98                    &0.99\\
    & DU2 Offset & keV                                & 0.06                   & 0.04                   & 0.02                    &-0.04\\
    & DU3 Slope &                                    & 0.98                    & 0.98                   & 0.98                    &0.97\\
    & DU3 Offset & keV                                & 0.01                   & 0.04                   & 0.01                    &0.03\\ \hline
  Test statistic & $\chi^2$/dof &                     & 567/477                & 616/480                & 524/476                 &591/475\\
  \hline

  \end{tabular}
  \begin{flushleft}
    Notes: \\
    The 90\% confidence intervals ($\Delta\chi^2=2.706$) for each parameter are noted above. \\
    $\ast$: The disk normalization is defined as Norm $=(R_{eff}/D_{10})^2*\cos \theta $, where $R_{eff}$ is the apparent inner radius of the accretion disk in km, $D_{10}$ is the distance to the source in units of 10~kpc, and $\theta$ is the inclination angle of the disk in degrees.  \\
    $\dagger $: The blackbody normalization is defined as Norm $=(R/D_{10})^2$, where $R$ is the radius of the emission region in km.\\
  \end{flushleft}
\end{table}

\begin{figure}[ht!]
  \centering
  \includegraphics[width=0.5\columnwidth]{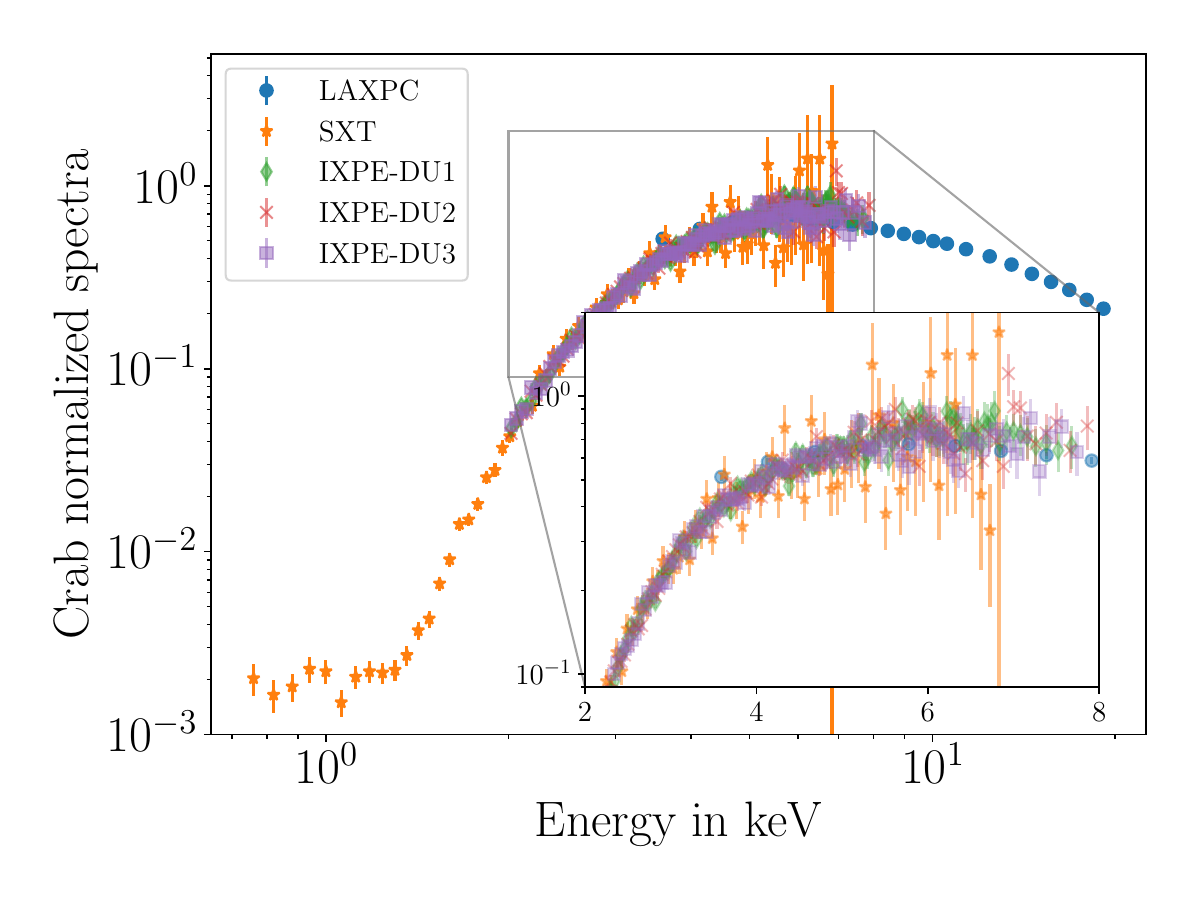}
  \caption{Crab normalized spectra extracted from one of the simultaneous intervals of \astr\ and \ixpe\ observations. The inset shows a zoom in view of the 2--8 keV range. The fluctuations in SXT spectra at higher energies is potentially due to low effective area and a lack of sufficient photons due to exclusion of a large inner region to prevent pile-up contamination.  }\label{fig:crab_norm_spec}
\end{figure}

As a sanity check, we removed the instrumental effects by normalizing the \src\ spectra with  Crab spectra extracted from a recent observation. 
In case of \astr, we use the Crab observation (observation id: AS1C09\_009T01\_9000006406) taken on August 22, 2024 while for \ixpe, we use the Crab observation (observation id: 03009601) taken on August 19, 2024. For extraction of the spectra, we follow the same procedure as outlined in section \ref{appsec:data_red} with only difference in the region selection for \astr-SXT (annular region of inner radius 13' and outer radius of 15'). On normalizing the \src\ spectra with Crab spectra and scaling to correct for effective area differences in the extraction method, we find that the data points align across the instruments (Figure~\ref{fig:crab_norm_spec}). As these are simultaneous intervals, the alignment indicates that discrepancies observed across the instruments are inherent to the instruments themselves and vanish on comparison with a standard source.

\end{appendix}





\end{document}